\pdfoutput=1
\documentclass[aps,twocolumn,superscriptaddress, prl]{revtex4-2}
\usepackage{graphicx}
\usepackage{dcolumn}
\usepackage{bm}
\usepackage{amsmath}
\usepackage{amssymb}
\usepackage{braket}
\usepackage[caption=false]{subfig}
\usepackage[colorlinks=true]{hyperref}
\usepackage{microtype}

\def\*#1{\mathbf{#1}}
\def\!#1{\mathbf{\hat#1}}
\def\mc#1{\mathcal{#1}}
\def\mb#1{\mathbb{#1}}

\def\t#1{\text{#1}}

\def\bs#1{\boldsymbol{#1}}
\def\grad{\boldsymbol{\nabla}}

\usepackage[draft,inline,nomargin]{fixme}

\fxsetup{theme=color,mode=multiuser}

\FXRegisterAuthor{sk}{ask}{\color{red}SK}
\FXRegisterAuthor{je}{aje}{\color{blue}JE}

\begin{document}

\preprint{APS/123-QED}

\title{Nematronics: Reciprocal coupling between ionic currents and nematic dynamics}

\author{Chau Dao}
    \affiliation{Department of Physics and Astronomy and Bhaumik Institute for Theoretical Physics, University of California, Los Angeles, California 90095, USA}
\author{Jeffrey C. Everts}
    \affiliation{Institute of Physical Chemistry, Polish Academy of Sciences, 01-224 Warsaw, Poland}
    \affiliation{Institute of Theoretical Physics, Faculty of Physics, University of Warsaw, Pasteura 5, 02-093 Warsaw, Poland}
\author{Miha Ravnik}
    \affiliation{Faculty of Mathematics and Physics, University of Ljubljana, Jadranska 19, 1000 Ljubljana, Slovenia}
    \affiliation{Department of Condensed Matter Physics, Jozef Stefan Institute, Jamova 39, 1000 Ljubljana, Slovenia}
\author{Yaroslav Tserkovnyak}
    \affiliation{Department of Physics and Astronomy and Bhaumik Institute for Theoretical Physics, University of California, Los Angeles, California 90095, USA}

\date{\today}

\begin{abstract}
Adopting a spintronics-inspired approach, we study the reciprocal coupling between ionic charge currents and nematic texture dynamics in a uniaxial nematic electrolyte. Assuming quenched fluid dynamics, we develop equations of motion analogously to spin torque and spin pumping. Based on the principle of least dissipation of energy, we derive the adiabatic ``nematic torque" exerted by ionic currents on the nematic director field as well as the reciprocal motive force on ions due to the orientational dynamics of the director. We discuss several simple examples that illustrate the potential functionality of this coupling. Furthermore, using our phenomenological framework, we propose a practical means to extract the coupling strength through impedance measurements on a nematic cell. Exploring further applications based on this physics could foster the development of nematronics -- nematic iontronics. 
\end{abstract}

\maketitle

\textit{Introduction.}|The reciprocal coupling between magnetic and electric degrees of freedom is the hallmark of spintronics \cite{Zutic2004,Tserkovnyak2018perspective}. There are myriad proposals to exploit electron mediated spin torque and spin pumping for novel devices and applications \cite{Wolf2001}, such as creating energy storage devices \cite{Vedmedenko2014,Jones2020}, spin-based memory \cite{Chappert2010,Parkin2008,diao:2007,Locatelli2014}, and methods for long-range signal transport \cite{Zutic2004,Cornelissen2015, Zou2020}. For soft condensed matter and biological systems, we extend the spintronics approach of studying effective torques and motive forces to nematic electrolytes (i.e. ion-doped nematic liquid crystal). Here, the nematic director plays the role of spin and ions take the place of electrons as the main charge carriers. Nematic electrolytes are natural systems to apply a spintronics-based approach because they share similar order parameter spaces and homotopic properties with magnetic systems \cite{Mermin1979,Prost1993,Vertogen1988,Nakahara2003,Lavrentovich2006}.
\begin{figure}
    \centering
    \includegraphics[width = \linewidth]{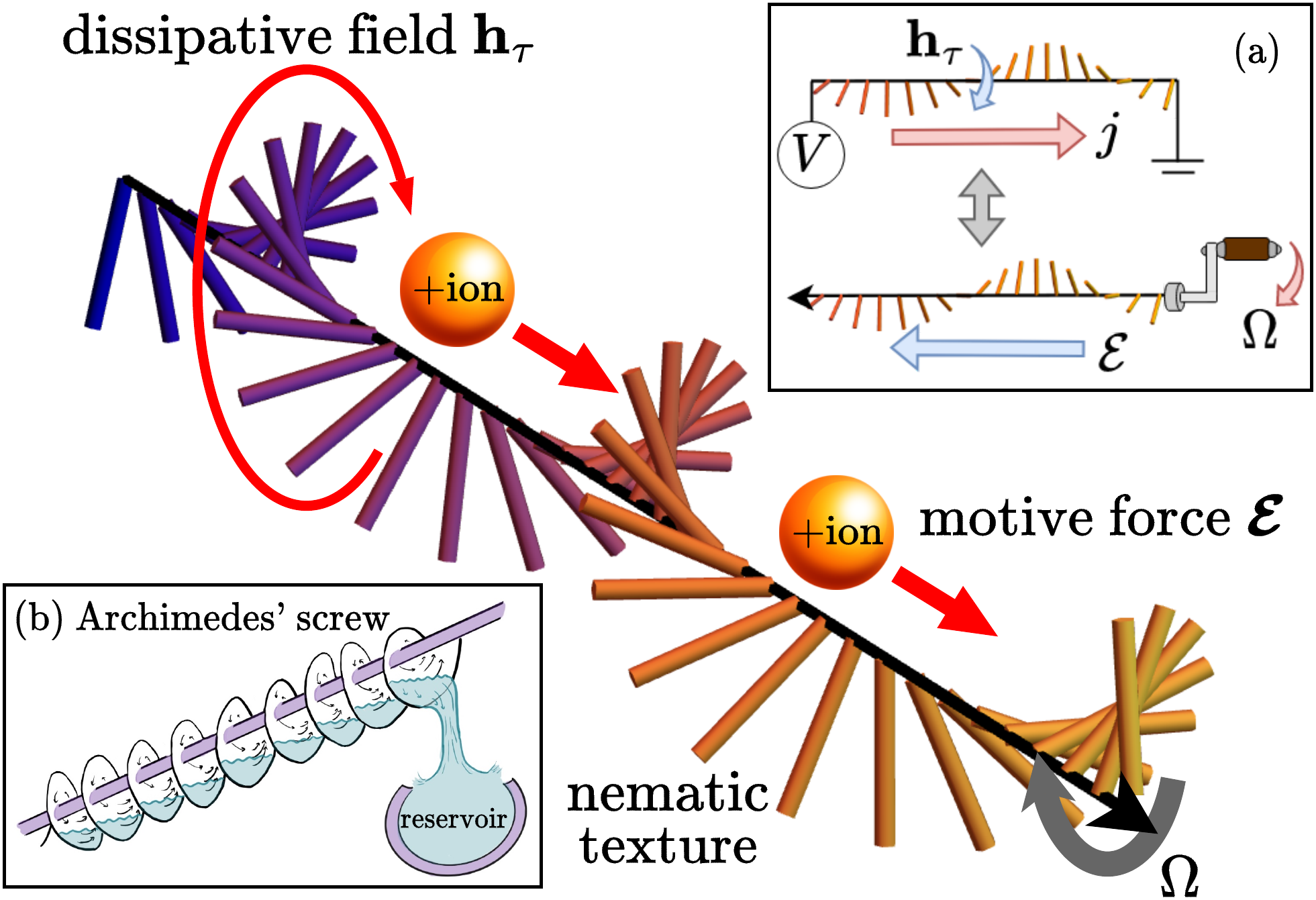}
    \caption{The nematic analog of the Archimedes' screw. An ionic current (driven by a voltage bias) induces an effective field $\*h_\tau$ on the nematic director field, causing the director field to rotate. Inversely, ions are pumped by the motive force $\bs{\mathcal{E}}$, which is induced when the director is rotated at frequency $\Omega$ through external means. In panel (a), red arrows denote the external drives and blue arrows denote induced forces. Panel (b) depicts the actual Archimedean screw transporting water, in analogy with the nematic screw transporting charge. The signs of $\*h_\tau$ and $\bs{\mc E}$ depend on the sign of $\gamma$, which has been chosen in this figure to mimic the actual screw.}
    \label{fig:ArchimedeanScrew}
\end{figure}

Drawing inspiration from spintronics, we will study a reciprocal coupling between ionic and nematic degrees of freedom, using it to drive nematic texture and pump charge current. This coupling is visualized in Fig. \ref{fig:ArchimedeanScrew} as the analog of Archimedes' screw, a hydraulic machine that can be used to either pump water or generate energy as a turbine \cite{archscrew}. Similarly, nematic dynamics driven by external means \cite{Koso:1988} pump ionic currents and, reciprocally, an ionic current (realized via an external electric field) exerts a torque on the nematic texture, generating work. Previous works on nemato-ionic interactions have focused on liquid-crystal enabled electrokinetics, in which applied electric fields are used to induce osmotic flows or to transport suspended particles within liquid crystals \cite{LavrentovichRev2020,Tovkach2017,Tovkach2016,Conklin2018} From a symmetry perspective, our proposed coupling is similar to the Lehmann effect, which describes the rotation of chiral nematics due to a temperature gradient \cite{lehmann1900,Poy:2019}, electric fields \cite{madhusudana:1989}, or concentration gradients \cite{brand2006, tabe2003}. Moreover, via the inverse Lehmann effect, a chiral nematic can pump particle flux \cite{Svensek2008}. For the Lehmann effect, the chirality of the system breaks inversion symmetry. This consequently leads to an effective field (and motive force satisfying Onsager reciprocity) stemming from the chiral coupling term \cite{Svensek2008}. In contrast, here we instead utilize couplings to spatial gradients of the nematic texture to reduce the symmetries of the director field configuration and study the subsequent dynamics. Such gradient field configurations can be induced by boundary conditions or topological defects.

In this Letter, we develop equations of motion analogously to spin torque and spin pumping for nematic electrolytes. To this end, we first specify the free energy and Rayleighan, then derive the dynamic equations from the principle of least dissipation. We discuss how, via this coupling, nematic dynamics give rise to inductance and how ionic current can be used to transport topological defects. By doing so, we aim at laying out building blocks for nematronics.

\textit{Dynamic equations.}|In the minimal model, consider a closed container filled with an incompressible dilute uniaxial nematic electrolyte. The orientational structure is described by the nematic director field $\* n(\*r)$, with unit norm. The fluid dynamics can be quenched when the fluid stiffness and finite system size results in faster fluid motion than director and ionic dynamics \cite{supmat}. The electrolyte consists of two freely moving monovalent ion species, with cationic density $\rho_+(\*r)$ and anionic density $\rho_-(\*r)$. Formally, this system can be decomposed into a charge sector, characterized by $\rho(\*r) = e[\rho_+(\*r) - \rho_-(\*r)]$ and a neutral ``osmotic" sector, characterized by $\Tilde{\rho}(\*r) = \rho_+(\* r) + \rho_-(\* r)$. Here, $e$ is the electric charge. Suppose we drive the system electrically so that the time scales of the two sectors separate. This results in faster plasmonic charge dynamics on top of the slower diffusive osmotic dynamics. Focusing on the faster charge sector, the remaining degrees of freedom are $\* n$ and $\rho$ \cite{osmotic}. Furthermore, the response in the charge sector may be approximated by the dynamics of the more mobile of the two ion species.

The equations of motion governing the nemato-ionic response are 
\begin{subequations}
\begin{align}
\label{eq:dtn}
    \partial_t \*n &= -\frac{1}{\alpha}(\*h_\perp + \*h_\tau), \\
\label{eq:ion_current}
    \*j &= -\beta e\rho_0\bs{\mc D}\cdot  \left(\bs\nabla \mu + \boldsymbol{\mathcal{E}}\right),\\
    \intertext{which are written in terms of Onsager-reciprocal constitutive relations}
\label{eq:htau}
    \*h_\tau &= \gamma\left(\*j\cdot\grad\right)\*n,\\
\label{eq:motive_force}
\mathcal{E}_i &= \gamma \,(\partial_i \*n)\cdot(\partial_t \*n).
\end{align}
\end{subequations}
Once the free energy $\mc F$ is specified, $\*h_\perp = \*n\times \delta_\*n \mc F \times \*n$ in Eq. (\ref{eq:dtn}) is the effective field thermodynamically conjugate to the director field $\*n$, with components parallel to $\*n$ projected out to fix $|\* n| = 1$. $\alpha$ is a phenomenological parameter characterizing the ``rotational viscosity" \cite{Zakharov1999} and $\beta = (k_{B}T)^{-1}$. $\bs{\mc D}$ is the effective diffusivity tensor in the charge sector, which accounts for the anisotropy of the nematic texture. It is constructed on symmetry grounds as $\mc D_{ij} = \mc D_\perp \delta_{ij} + \Delta \mc D \, n_i n_j$. $\rho_0$ is the homogenous bulk ion density. $\mu = \delta_\rho \mc F$ is the effective electrochemical potential conjugate to the charge density $\rho$. Finally, $\gamma$ is a phenomenological coefficient parameterizing the rotation imparted upon the nematic texture due to $\*j$.

The ion-induced field $\*h_\tau$ in Eq. (\ref{eq:dtn}) is fully analogous to the adiabatic ``spin-transfer torque" in spin systems \cite{Ralph2008,Tserkovnyak2002}, and describes the torque exerted by the electric charge current $\*j$ onto the nematic texture. Reciprocally, director dynamics induces a motive force ${\bs{\mc E}}$, given by Eq. (\ref{eq:motive_force}), which pumps a diffusive charge current $\*j$, given by Eq. (\ref{eq:ion_current}). $\*h_\tau$ and $\bs{\mc E}$ are only nonzero for systems out of equilibrium, i.e. displaced from the minimum free energy. We note that $\*h_\tau$ and $\bs{\mc E}$ are independently symmetric under rotations in real and order parameter space. This symmetry is artificial because, in general, it is broken by additional terms. Moreover, only the combined symmetry corresponding to simultaneous rotations of real space and director orientation survives in a fluid. The Supplemental Material provides a more complete discussion of these ``artificial" symmetries \cite{supmat}.

\textit{Framework.}|With the free energy $\mc F$ and Rayleighan $\mc R$, we employ the principle of least dissipation of energy, $\delta_{\dot{q_i}}\left\{\partial_t\mc F + \int d^3\*r \, \mc R\right\} = 0$, to derive the dynamic equations. In this expression, $\dot{q}_i$ are the generalized velocities $\partial_t\*n$ and $\*j$. We have assumed that the system has a uniform nematic order parameter. However, it is straightforward to generalize by constructing $\mc F$ and $\mc R$ with the tensorial nematic order parameter $\*Q$ instead of $\*n$ \cite{Qian1998,Tovkach2017}. Furthermore, the material flow of the nematic fluid is not an explicit degree of freedom. This can be a limiting assumption in electrohydrodynamic systems \cite{Tovkach2016,Tovkach2017,Conklin2018} but qualitatively accepted, for example, in confined geometries \cite{Chandragiri2020,HernandezOrtiz2011}. 

Our model incorporates ionic degrees of freedom into the Ericksen-Leslie formalism of nematodynamics \cite{Leslie1979}. Similar models have been derived which focused on electrohydrodynamic effects but did not include the reciprocal nemato-ionic coupling \cite{Tovkach2016,Tovkach2017,Calderer2016}. In our model, the total free energy is
\begin{equation}
\begin{aligned}\label{eq:free_energy}
    \mc F[\*n, \rho, \Tilde{\rho},\psi]=& \int d^3\*r\, \bigg\{ \frac{K}{2}(\partial_in_j)^2 +F_{\t{ion}}(\rho,\Tilde{\rho})\\&+ \rho\psi - \frac{1}{8\pi}\epsilon_{ij}(\partial_i\psi)(\partial_j\psi)\bigg\}.
\end{aligned}
\end{equation}
$K$ is the elastic constant, $\psi$ is the electric potential, and $\epsilon_{ij}=\epsilon_\perp\delta_{ij}+\Delta\epsilon\,n_in_j$ is the dielectric tensor constructed analogously to the diffusivity tensor.

The first term is the free energy density of the nematic texture which accounts for the elasticity of the liquid crystal. We assume the ``equal constant approximation," where splay, twist, and bend modes have equal elastic constant $K$ \cite{doi:10.1080/15421406.2017.1289425}. This corresponds to the aforementioned ``artificial" symmetries in which the system is separately isotropic in real and order parameter space. The second term, $F_\t{ion}$, describes the nonelectrostatic ionic part of the free energy density. The final two terms are the electrostatic contributions to the free energy. They give the Poisson equation in an anisotropic dielectric upon $\delta_\psi\mathcal{F}=0$,
\begin{equation}
\bs\nabla\cdot\left[\boldsymbol{\epsilon}\cdot\bs\nabla\psi\right]=-4\pi\rho \label{eq:Poisson}.
\end{equation}
Note that the flexoelectric free energy could be included, but our focus is on the kinematic effects corresponding to the nemato-ionic response \cite{supmat,chandrasekhar_1992,Shen2010}.

The Rayleigh functions capturing linear dissipative forces are positive definite and quadratic in generalized velocities. They are given by
\begin{subequations}
\begin{align}
\label{eq:rayleigh_nn}
\mc R_{nn} &= \frac{\alpha}{2}(\partial_t\*n)^2, \\
\label{eq:rayleigh_uu}
    \mc R_{jj} &= \frac{k_BT}{2e\rho_0} \, \*j\cdot {\bs{\mc D}}^{-1} \cdot \*j,\\
\label{eq:rayleigh_nu}
    \mc R_{nj} &= \gamma \, \partial_t\*n\cdot[(\*j\cdot\bs\nabla)\*n].
\end{align}
\end{subequations}
Here, $\mc R_{nn}$ governs the relaxation of the director field to its equilibrium configuration. $\mc R_{jj}$ describes the friction of electric charge currents, where the effective diffusivity tensor is $\bs{\mc D}$. Eqs (\ref{eq:rayleigh_nn}) and (\ref{eq:rayleigh_uu}) can be found in Refs. \cite{Tovkach2016,Tovkach2017,Conklin2018}. $\mc R_{nj}$ is the friction term coupling electric charge current to director dynamics. It respects the previously discussed ``artificial" symmetries and is constructed to leading order in spatial derivatives of $\*n$. $\mc R$ being positive-definite constrains $\bs{\mc D}$, $\alpha$, and $\gamma$ \cite{Gyarmati1970, Sonnet:2004}. 

Having specified $\mc F$ and $\mc R$, we now apply the principle of least dissipation. Taking the functional derivative of $\partial_t\mc F + \int d^3\*r \, \mc R$ with respect to $\partial_t \*n$ yields Eq. (\ref{eq:dtn}), describing time dynamics of the director. Varying with respect to the charge current $\*j$ alongside invoking the continuity equation, $\partial_t \rho = - \bs\nabla \cdot \*j$, yields Eq. (\ref{eq:ion_current}). For a dilute electrolyte, $F_\t{ion}$ is the sum of the entropic ideal gas contributions of each ionic species. When deviations of $\rho_\pm$ from the homogenous bulk ion density $\rho_0$ are small, expanding $\mu$ in Eq. (\ref{eq:ion_current}) about $\rho_0$ yields $\mu = \psi + \rho/\chi$, up to a constant. $\chi = 2e^2\rho_0/k_BT$ is the effective charge compressibility. Furthermore, for this $F_\t{ion}$, $\bs{\mc D}$ in Eq. (\ref{eq:ion_current}) is the sum of the diffusivity tensors ${\bs{\mc D}}_{\pm}$ of each ionic species. Eqs. (\ref{eq:htau}) and (\ref{eq:motive_force}) are obtained by taking the variational derivative of $\mc R_{nj}$ with respect to $\partial_t\*n$ and $\*j$, respectively. The two equations follow Onsager reciprocity, which could have been invoked in lieu of Rayleigh functions and the principle of least dissipation \cite{Onsager1931,Gyarmati1970}. To illustrate the applications of these effects, we first study a simple example in which a nontrivial nematic texture induced by strong anchoring is electrically driven out of equilibrium.

\begin{figure}
    \centering
    \includegraphics[width = \linewidth]{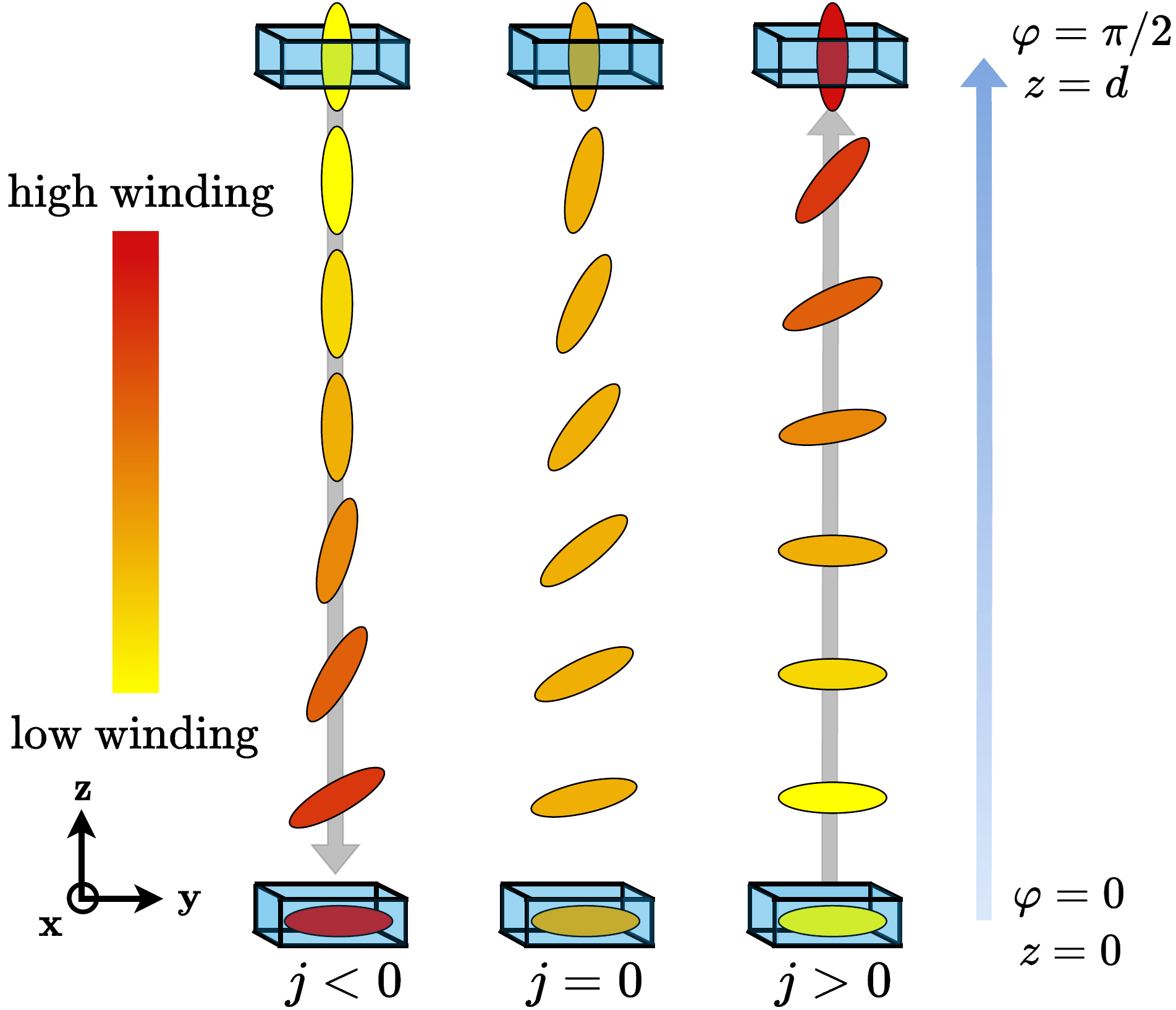}
    \caption{Nematic texture for different current values. Three cases are shown for a nematic slab between two electrodes. The fixed boundary conditions are $\*n(d) = \*z$ and $\*n(0) = \*y$, signified by the nematogens enclosed in boxes. The director winds $\Delta\varphi = \pi/2$ from the bottom to the top plate. The leftmost slab is biased with $j<0$, the middle unbiased, and the rightmost with $j>0$. The nematogens are colored according to a winding heat map. The winding is localized in the direction of the charge current, with winding density decay length $\xi = K/\gamma j$.}
    \label{fig:nematic_switch}
\end{figure}

\textit{Nematic-induced inductance.}|Let us consider a slab of nematic electrolytic fluid uniform in the $xy$ plane and of thickness $d$ in the $z$ direction. We impose surface boundary conditions $\*n (d)= \*z$ and $\*n (0)= \*y$. We define $\varphi(z)$ as the angle of $\*n_\|$, the planar projection of $\*n$ onto the $yz$ easy-plane, relative to the $y$ axis. To satisfy the boundary conditions, the change in $\varphi$ from the bottom to top plate is $\Delta\varphi = (n +1/2)\pi$ for $n\in \mb Z$. Fig. \ref{fig:nematic_switch} depicts our proposed setup for the simplest case of $n=0$, known also as the hybrid-aligned nematic cell \cite{Matsumoto1976}. 

Suppose we bias the system with an alternating current given by $\* j(t)=j(t)\*z$.
For a sufficiently slowly varying current $j$, the solution for the nematic texture is well-approximated by the quasistatic solution. We consequently set $\partial_t \*n=0$ in Eq. (\ref{eq:dtn}) and parameterize the nematic texture in terms of the winding density in the $yz$ plane, $\eta(z) \equiv (\*n \times \partial_z \*n) \cdot \*x$. Focusing on the linear response, electrostatic contributions to $\*h_\perp$ are not included since they are quadratic in $\*E$. For a full detailed analysis, the dielectric term $\Delta\epsilon(\*n\cdot\*E)^2$
would be included \cite{chandrasekhar_1992,Frisken:1989}. However, the effects due to the nemato-ionic coupling can always be qualitatively distinguished from e.g., the Freedericksz effect, since the former depends on the sign of $\*E$ while the latter does not. Equation (\ref{eq:dtn}) then governs the winding density as $K\, \partial_z\eta = \gamma j\eta$. The coupling of current to the winding rather than to nematic orientation further distinguishes this effect from the Freedericksz effect. 

Integrating Eq. (\ref{eq:dtn}), the quasistatic solution for the winding density is
\begin{equation}\label{eq:eta}
    \eta(z) = \frac{1}{\xi}\frac{e^{z/\xi}}{e^{d/\xi}-1}\left(n\pi+\frac{\pi}{2}\right),
\end{equation}
where $\xi = K/\gamma j$ is the length scale at which the winding density decays. Winding with $n=0,-1$ are stable ground states. All other values of $n$ are metastable states since the winding can be smoothly unwound in multiples of $2\pi$ by allowing the director to rotate out of the $yz$ plane. These states can be stabilized by including easy-plane anisotropy, disallowing out-of-plane rotations. Furthermore, these metastable states can be made accessible by utilizing chiral liquid crystals. The winding density is dragged along the direction of the current flow. If $|j| \gg K/\gamma d$, the winding density localizes at $z=d$ or $z=0$, for positive and negative values of $j$ respectively. In these cases, the nematic texture switches between configurations that are parallel or orthogonal to the $xy$ plane.

The backaction by the nematic dynamics on the electrical response induces the motive force given in Eq. (\ref{eq:motive_force}). In this setup, $\mathcal{E}_z = \gamma (\partial_z \varphi)(\partial_t \varphi)$, where the angle $\varphi(z,t)$ is obtained by integrating Eq. (\ref{eq:eta}) while requiring $|\*n_\|| = 1$. $\partial_t\varphi$ can be formally understood as the winding flux. The motive force $\boldsymbol{\mc E}$,  which can be understood as a fictitious electric field, induces an effective potential drop between the bottom and top plates, in addition to the actual electric field $\*E = -\bs\nabla\psi$. Integrating the motive force $\bs{\mc E}$ over the thickness $d$ and taking the leading term in $j$, the instantaneous effective potential difference is
\begin{equation}
    \Delta V(t) = \left(n\pi+\frac{\pi}{2}\right)^2\frac{\gamma^2d}{12K}(\partial_tj).
\end{equation}
Fourier transforming into the frequency domain, the corresponding induced impedance of the nematic slab is
\begin{equation}
    Z(\omega) \equiv \frac{\Delta V(\omega)}{j(\omega)}= i\omega\left(n\pi+\frac{\pi}{2}\right)^2\frac{\gamma^2d}{12K},
\end{equation}
which has the inductive form \cite{impedance}. Suppose a capacitor is filled with a slab of nematic electrolyte. Driving the capacitor with an alternating current and measuring the impedance yields the strength of the reciprocal coupling $\gamma$ relative to elasticity $K$. Moreover, as the integrated winding increases, so would the measured effective inductance.

\begin{figure}
    \centering
    \includegraphics[width = \linewidth]{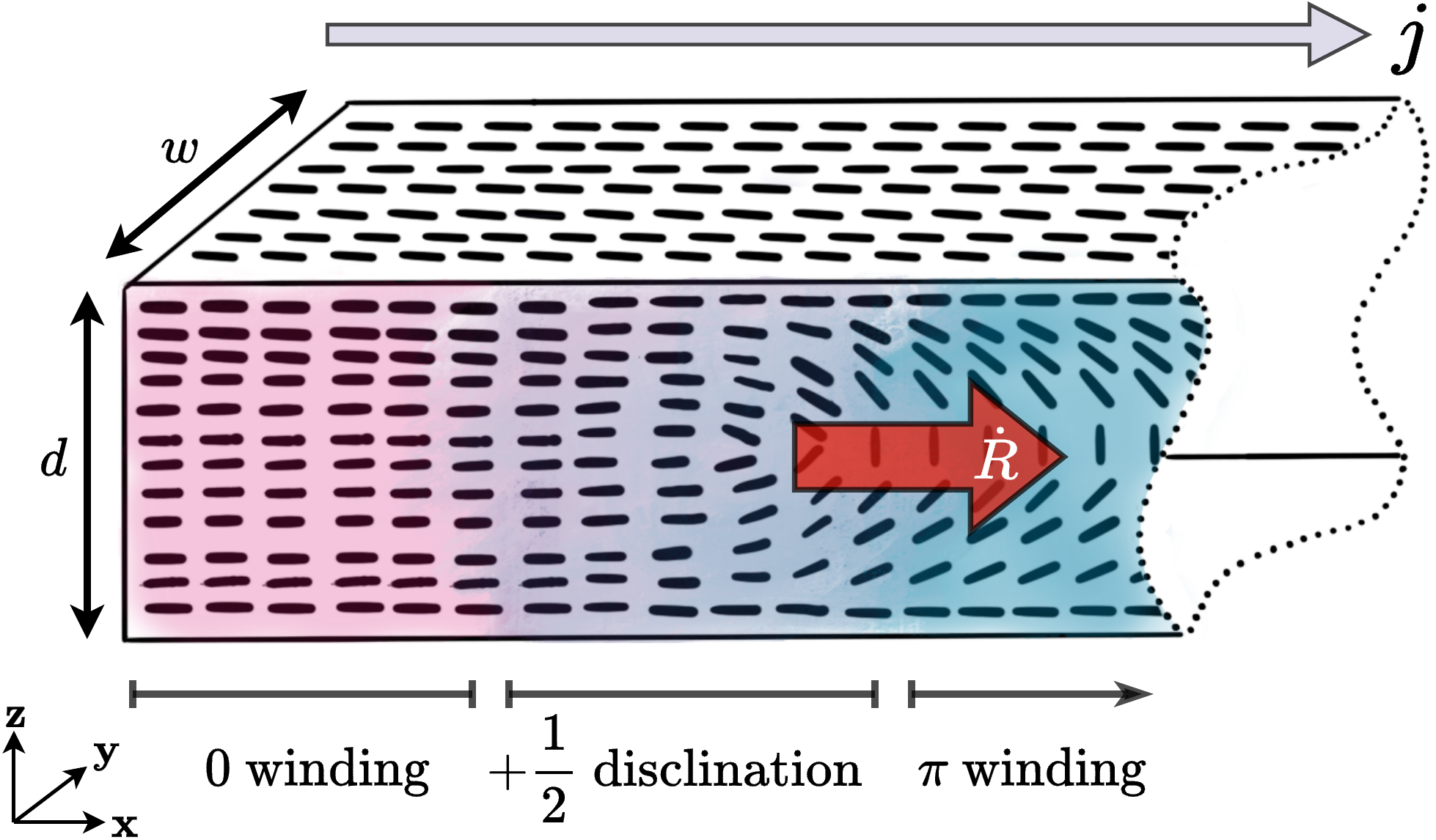}
    \caption{Setup of current driven nematic defect. The nematic slab has a pink domain of $0$ winding, a blue domain of $\pi$ winding, and a disclination region of size $d$ transitioning between pink and blue. The texture on the top, bottom, and left faces is fixed to point along the $x$ direction. The defect is shown to move in the $x$ direction with velocity $\dot R$. The velocity can be increased when an electric current $j$ is applied.}
    \label{fig:soliton}
\end{figure}

\textit{Line disclination dynamics.}|In addition to boundary conditions, topological defects also imprint nontrivial nematic textures. This leads to dynamic effects when the system is subjected to an electrical current flow. As a minimal example, we study the response of a line disclination with a planar cross-section with winding number $+1/2$ to an electrical current. Consider a nematic slab uniform in the $y$ direction with a width of $w$, of thickness $d$ in the $z$ direction, and length in the $x$ direction larger than $d$. On the top, bottom, and left-most face, the director is strongly anchored to point along the $x$ direction. The setup is depicted in Fig. \ref{fig:soliton}. 

Suppose there are two domains: a small domain with 0 winding from the bottom to the top face and a large domain in which the director uniformly undergoes $\pi$ winding. Transitioning from one domain to the other, we find a disclination of radius $d$, with a line singularity along the $y$ axis. This setup has been experimentally realized by Sandford O'Neill \textit{et al.} \cite{Sandford2020}. However, their focus was on controlling the defect by utilizing the electric coupling stemming from $(\*n\cdot\*E)^2$ anisotropy, rather than the nemato-ionic coupling.

We will study the forces on the defect, as well as derive its terminal velocity. Approximating the defect as a ``rigidly moving soliton," we employ the ansatz $\*n(\*r,t)\rightarrow \*n_0(\*r + \*R(t))$ \cite{Thiele1973}. In doing so, we place the time dependence of the director field into $\*R(t)$, the position of the defect core. The force on the defect is $F_s = -\partial \mc F/\partial R - \partial\mc R/\partial \dot R$ for free energy $\mc F$ and Rayleighan $\mc R$. In the vicinity of the disclination, the angle of the projection of $\*n$ to the $xz$ plane is $\varphi(\*r) = (\phi+\pi)/2$. Here, $\phi$ is the polar angle relative to the $x$ axis \cite{Imura1973}. Away from the disclination, the nematic texture will distort to match the boundary conditions. 

Using the Rayleighans in Eqs. \eqref{eq:rayleigh_uu}-\eqref{eq:rayleigh_nu} and the free energy in Eq. (\ref{eq:free_energy}), the force on the defect is
\begin{equation}
   F_s = \frac{Kw\pi^2}{2d} - \frac{\alpha w\pi}{4}\left(\dot R+ \frac{\gamma j}{\alpha}\right) \ln\left(\frac{d}{2r_c}\right),
\end{equation}
with $r_{c}$ the defect core size. The first term is a constant force due to the elastic relaxation of the nematic texture. To lower the free energy, the 0 winding domain will grow while the $\pi$ winding domain shrinks. The second term is a drag force on the disclination and stems from $\mc R_{nn}$ and $\mc R_{nj}$. The dissipation function $\mc R_{nn}$ results in a core velocity-dependent drag force. $\mc R_{nj}$, the friction between charge current and nematic texture, uses current $j$ as a handle to apply a force. Combining the Rayleighans, we find an effective Rayleighan to leading order in $j$,
\begin{equation}
   \mc R_\t{eff} = \frac{\alpha w\pi}{8}\left(\dot R + \frac{\gamma j}{\alpha}\right)^2\ln\left(\frac{d}{2r_c}\right).
\end{equation}
$\mc R_\t{eff}$ describes the solitonic viscosity, with damping parameter $\alpha w\pi\ln(d/2r_c)/8$. The generalized velocity is $\dot R + \gamma j/\alpha$, the defect velocity modified by the ionic current. Instead of using $\partial_t\*n$ and $\*j$ as the generalized velocities, we could instead use this collective degree of freedom in the Rayleighan.

The terminal velocity of the disclination is 
 \begin{equation}
     \dot R_t = \frac{2\pi K}{\alpha d\ln(d/2r_c)} - \frac{\gamma j}{\alpha}.
 \end{equation}
When $\gamma = 0$, $\dot R_t$ agrees with previous studies for the velocity of a disclination in the absence of fluid motion \cite{Tang2019,Imura1973}. Measurement of $\dot R_{t}(j)$ can provide $\gamma/\alpha$, parameterizing the relative strengths of the nemato-ionic coupling to the damping $\alpha$. The motion of the disclination can be understood as topological texture electrophoresis, in which topological textures are transported in the absence of fluid motion or suspended particles. Electric current provides a robust handle to manipulate topological defects, enabling technological applications such as microparticle transport and spatial light modulators \cite{Yoshida2015,Nassiri2018}. 

\textit{Discussion.}|Thus far, we have constructed the scaffolding for future studies of dynamic phenomena as well as applications in nematronics. Moving forward, we can consider nematronic devices which exploit this coupling, with a focus on topological aspects. For example, building up winding via manipulating topological defects could lead to energy storage devices \cite{Patil2020,Tserkovnyak2018}. Current driven transport of nematic defects for signal transport would also be an intriguing avenue of research. Nematic hedgehogs and disclinations are nonlocal topological objects \cite{Nakahara2003}, which makes them promising candidates for information carriers in soft matter systems \cite{Kos2022}.

From a theoretical perspective, studying these effects in different liquid crystalline phases, e.g. twist-bend nematic phases \cite{Selinger:2018} and blue phases \cite{Yeomans:2007,Castles2010}, is a  potentially interesting research direction. To elucidate the origins of $\gamma$, one could study the coupling microscopically rooted in interactions between ionic charge fluctuations and nematogen polarizability. In spintronics, research into driving ferromagnetic domain wall motion with spin-polarized electrical current has unraveled novel physics \cite{Berger1978,Parkin2008,Kittel1949}. It could be fruitful to proceed along a similar line by studying the electrical response of domain walls in the nematic and ferroelectric nematic phase \cite{Mertelj2020, Lavrentovich2020, Sandford2020}. Additionally, studies on isotropic electrolytes show that solutions with multiple ion species with different valencies \cite{Warren:2020} and charged colloidal particles can lead to intriguing effects \cite{Avni:2019} that could be also studied in nematic systems. 

\begin{acknowledgments}
\textit{Acknowledgments. } This work was primarily supported by the U.S. Department of Energy, Office of Basic Energy Sciences under Grant No. DE-SC0012190 (C.D. and Y.T.). J.C.E. acknowledges financial support from the Polish National Agency for Academic Exchange (NAWA) under the Ulam programme Grant No. PPN/ULM/2019/1/00257. M.R. acknowledges funding from Slovenian research agency ARRS grants  P1-0099, N1-0195, and J1-2462, and EU ERC AdG LOGOS.
\end{acknowledgments}

\bibliography{ms.bib}

\begin{thebibliography}{65}%
\makeatletter
\providecommand \@ifxundefined [1]{%
 \@ifx{#1\undefined}
}%
\providecommand \@ifnum [1]{%
 \ifnum #1\expandafter \@firstoftwo
 \else \expandafter \@secondoftwo
 \fi
}%
\providecommand \@ifx [1]{%
 \ifx #1\expandafter \@firstoftwo
 \else \expandafter \@secondoftwo
 \fi
}%
\providecommand \natexlab [1]{#1}%
\providecommand \enquote  [1]{``#1''}%
\providecommand \bibnamefont  [1]{#1}%
\providecommand \bibfnamefont [1]{#1}%
\providecommand \citenamefont [1]{#1}%
\providecommand \href@noop [0]{\@secondoftwo}%
\providecommand \href [0]{\begingroup \@sanitize@url \@href}%
\providecommand \@href[1]{\@@startlink{#1}\@@href}%
\providecommand \@@href[1]{\endgroup#1\@@endlink}%
\providecommand \@sanitize@url [0]{\catcode `\\12\catcode `\$12\catcode
  `\&12\catcode `\#12\catcode `\^12\catcode `\_12\catcode `\%12\relax}%
\providecommand \@@startlink[1]{}%
\providecommand \@@endlink[0]{}%
\providecommand \url  [0]{\begingroup\@sanitize@url \@url }%
\providecommand \@url [1]{\endgroup\@href {#1}{\urlprefix }}%
\providecommand \urlprefix  [0]{URL }%
\providecommand \Eprint [0]{\href }%
\providecommand \doibase [0]{https://doi.org/}%
\providecommand \selectlanguage [0]{\@gobble}%
\providecommand \bibinfo  [0]{\@secondoftwo}%
\providecommand \bibfield  [0]{\@secondoftwo}%
\providecommand \translation [1]{[#1]}%
\providecommand \BibitemOpen [0]{}%
\providecommand \bibitemStop [0]{}%
\providecommand \bibitemNoStop [0]{.\EOS\space}%
\providecommand \EOS [0]{\spacefactor3000\relax}%
\providecommand \BibitemShut  [1]{\csname bibitem#1\endcsname}%
\let\auto@bib@innerbib\@empty
\bibitem [{\citenamefont {\ifmmode \check{Z}\else
  \v{Z}\fi{}uti\ifmmode~\acute{c}\else \'{c}\fi{}}\ \emph
  {et~al.}(2004)\citenamefont {\ifmmode \check{Z}\else
  \v{Z}\fi{}uti\ifmmode~\acute{c}\else \'{c}\fi{}}, \citenamefont {Fabian},\
  and\ \citenamefont {Das~Sarma}}]{Zutic2004}%
  \BibitemOpen
  \bibfield  {author} {\bibinfo {author} {\bibfnamefont {I.}~\bibnamefont
  {\ifmmode \check{Z}\else \v{Z}\fi{}uti\ifmmode~\acute{c}\else \'{c}\fi{}}},
  \bibinfo {author} {\bibfnamefont {J.}~\bibnamefont {Fabian}},\ and\ \bibinfo
  {author} {\bibfnamefont {S.}~\bibnamefont {Das~Sarma}},\ }\bibfield  {title}
  {\bibinfo {title} {Spintronics: Fundamentals and applications},\ }\href
  {https://link.aps.org/doi/10.1103/RevModPhys.76.323} {\bibfield  {journal}
  {\bibinfo  {journal} {Rev. Mod. Phys.}\ }\textbf {\bibinfo {volume} {76}},\
  \bibinfo {pages} {323} (\bibinfo {year} {2004})}\BibitemShut {NoStop}%
\bibitem [{\citenamefont {Tserkovnyak}(2018)}]{Tserkovnyak2018perspective}%
  \BibitemOpen
  \bibfield  {author} {\bibinfo {author} {\bibfnamefont {Y.}~\bibnamefont
  {Tserkovnyak}},\ }\bibfield  {title} {\bibinfo {title} {Perspective: (beyond)
  spin transport in insulators},\ }\href {https://doi.org/10.1063/1.5054123}
  {\bibfield  {journal} {\bibinfo  {journal} {J. Appl. Phys.}\ }\textbf
  {\bibinfo {volume} {124}},\ \bibinfo {pages} {190901} (\bibinfo {year}
  {2018})}\BibitemShut {NoStop}%
\bibitem [{\citenamefont {Wolf}\ \emph {et~al.}(2001)\citenamefont {Wolf},
  \citenamefont {Awschalom}, \citenamefont {Buhrman}, \citenamefont {Daughton},
  \citenamefont {von Molnár}, \citenamefont {Roukes}, \citenamefont
  {Chtchelkanova},\ and\ \citenamefont {Treger}}]{Wolf2001}%
  \BibitemOpen
  \bibfield  {author} {\bibinfo {author} {\bibfnamefont {S.~A.}\ \bibnamefont
  {Wolf}}, \bibinfo {author} {\bibfnamefont {D.~D.}\ \bibnamefont {Awschalom}},
  \bibinfo {author} {\bibfnamefont {R.~A.}\ \bibnamefont {Buhrman}}, \bibinfo
  {author} {\bibfnamefont {J.~M.}\ \bibnamefont {Daughton}}, \bibinfo {author}
  {\bibfnamefont {S.}~\bibnamefont {von Molnár}}, \bibinfo {author}
  {\bibfnamefont {M.~L.}\ \bibnamefont {Roukes}}, \bibinfo {author}
  {\bibfnamefont {A.~Y.}\ \bibnamefont {Chtchelkanova}},\ and\ \bibinfo
  {author} {\bibfnamefont {D.~M.}\ \bibnamefont {Treger}},\ }\bibfield  {title}
  {\bibinfo {title} {Spintronics: A spin-based electronics vision for the
  future},\ }\href {https://www.science.org/doi/abs/10.1126/science.1065389}
  {\bibfield  {journal} {\bibinfo  {journal} {Science}\ }\textbf {\bibinfo
  {volume} {294}},\ \bibinfo {pages} {1488} (\bibinfo {year}
  {2001})}\BibitemShut {NoStop}%
\bibitem [{\citenamefont {Vedmedenko}\ and\ \citenamefont
  {Altwein}(2014)}]{Vedmedenko2014}%
  \BibitemOpen
  \bibfield  {author} {\bibinfo {author} {\bibfnamefont {E.~Y.}\ \bibnamefont
  {Vedmedenko}}\ and\ \bibinfo {author} {\bibfnamefont {D.}~\bibnamefont
  {Altwein}},\ }\bibfield  {title} {\bibinfo {title} {Topologically protected
  magnetic helix for all-spin-based applications},\ }\href
  {https://doi.org/10.1103/PhysRevLett.112.017206} {\bibfield  {journal}
  {\bibinfo  {journal} {Phys. Rev. Lett.}\ }\textbf {\bibinfo {volume} {112}},\
  \bibinfo {pages} {017206} (\bibinfo {year} {2014})}\BibitemShut {NoStop}%
\bibitem [{\citenamefont {Jones}\ \emph {et~al.}(2020)\citenamefont {Jones},
  \citenamefont {Zou}, \citenamefont {Zhang},\ and\ \citenamefont
  {Tserkovnyak}}]{Jones2020}%
  \BibitemOpen
  \bibfield  {author} {\bibinfo {author} {\bibfnamefont {D.}~\bibnamefont
  {Jones}}, \bibinfo {author} {\bibfnamefont {J.}~\bibnamefont {Zou}}, \bibinfo
  {author} {\bibfnamefont {S.}~\bibnamefont {Zhang}},\ and\ \bibinfo {author}
  {\bibfnamefont {Y.}~\bibnamefont {Tserkovnyak}},\ }\bibfield  {title}
  {\bibinfo {title} {Energy storage in magnetic textures driven by vorticity
  flow},\ }\href {https://link.aps.org/doi/10.1103/PhysRevB.102.140411}
  {\bibfield  {journal} {\bibinfo  {journal} {Phys. Rev. B}\ }\textbf {\bibinfo
  {volume} {102}},\ \bibinfo {pages} {140411(R)} (\bibinfo {year}
  {2020})}\BibitemShut {NoStop}%
\bibitem [{\citenamefont {Chappert}\ \emph {et~al.}(2007)\citenamefont
  {Chappert}, \citenamefont {Fert},\ and\ \citenamefont
  {Van~Dau}}]{Chappert2010}%
  \BibitemOpen
  \bibfield  {author} {\bibinfo {author} {\bibfnamefont {C.}~\bibnamefont
  {Chappert}}, \bibinfo {author} {\bibfnamefont {A.}~\bibnamefont {Fert}},\
  and\ \bibinfo {author} {\bibfnamefont {F.~N.}\ \bibnamefont {Van~Dau}},\
  }\bibfield  {title} {\bibinfo {title} {The emergence of spin electronics in
  data storage},\ }\href {https://www.nature.com/articles/nmat2024} {\bibfield
  {journal} {\bibinfo  {journal} {Nat. Mater.}\ }\textbf {\bibinfo {volume}
  {6}},\ \bibinfo {pages} {813} (\bibinfo {year} {2007})}\BibitemShut {NoStop}%
\bibitem [{\citenamefont {Parkin}\ \emph {et~al.}(2008)\citenamefont {Parkin},
  \citenamefont {Hayashi},\ and\ \citenamefont {Thomas}}]{Parkin2008}%
  \BibitemOpen
  \bibfield  {author} {\bibinfo {author} {\bibfnamefont {S.~S.~P.}\
  \bibnamefont {Parkin}}, \bibinfo {author} {\bibfnamefont {M.}~\bibnamefont
  {Hayashi}},\ and\ \bibinfo {author} {\bibfnamefont {L.}~\bibnamefont
  {Thomas}},\ }\bibfield  {title} {\bibinfo {title} {Magnetic domain-wall
  racetrack memory},\ }\href
  {https://www.science.org/doi/abs/10.1126/science.1145799} {\bibfield
  {journal} {\bibinfo  {journal} {Science}\ }\textbf {\bibinfo {volume}
  {320}},\ \bibinfo {pages} {190} (\bibinfo {year} {2008})}\BibitemShut
  {NoStop}%
\bibitem [{\citenamefont {Diao}\ \emph {et~al.}(2007)\citenamefont {Diao},
  \citenamefont {Li}, \citenamefont {Wang}, \citenamefont {Ding}, \citenamefont
  {Panchula}, \citenamefont {Chen}, \citenamefont {Wang},\ and\ \citenamefont
  {Huai}}]{diao:2007}%
  \BibitemOpen
  \bibfield  {author} {\bibinfo {author} {\bibfnamefont {Z.}~\bibnamefont
  {Diao}}, \bibinfo {author} {\bibfnamefont {Z.}~\bibnamefont {Li}}, \bibinfo
  {author} {\bibfnamefont {S.}~\bibnamefont {Wang}}, \bibinfo {author}
  {\bibfnamefont {Y.}~\bibnamefont {Ding}}, \bibinfo {author} {\bibfnamefont
  {A.}~\bibnamefont {Panchula}}, \bibinfo {author} {\bibfnamefont
  {E.}~\bibnamefont {Chen}}, \bibinfo {author} {\bibfnamefont {L.-C.}\
  \bibnamefont {Wang}},\ and\ \bibinfo {author} {\bibfnamefont
  {Y.}~\bibnamefont {Huai}},\ }\bibfield  {title} {\bibinfo {title}
  {Spin-transfer torque switching in magnetic tunnel junctions and
  spin-transfer torque random access memory},\ }\href
  {https://doi.org/10.1088/0953-8984/19/16/165209} {\bibfield  {journal}
  {\bibinfo  {journal} {J. Phys. Condens. Matter}\ }\textbf {\bibinfo {volume}
  {19}},\ \bibinfo {pages} {165209} (\bibinfo {year} {2007})}\BibitemShut
  {NoStop}%
\bibitem [{\citenamefont {Locatelli}\ \emph {et~al.}(2014)\citenamefont
  {Locatelli}, \citenamefont {Cros},\ and\ \citenamefont
  {Grollier}}]{Locatelli2014}%
  \BibitemOpen
  \bibfield  {author} {\bibinfo {author} {\bibfnamefont {N.}~\bibnamefont
  {Locatelli}}, \bibinfo {author} {\bibfnamefont {V.}~\bibnamefont {Cros}},\
  and\ \bibinfo {author} {\bibfnamefont {J.}~\bibnamefont {Grollier}},\
  }\bibfield  {title} {\bibinfo {title} {Spin-torque building blocks},\ }\href
  {https://www.nature.com/articles/nmat3823} {\bibfield  {journal} {\bibinfo
  {journal} {Nat. Mater.}\ }\textbf {\bibinfo {volume} {13}},\ \bibinfo {pages}
  {11} (\bibinfo {year} {2014})}\BibitemShut {NoStop}%
\bibitem [{\citenamefont {Cornelissen}\ \emph {et~al.}(2015)\citenamefont
  {Cornelissen}, \citenamefont {Liu}, \citenamefont {Duine}, \citenamefont
  {Youssef},\ and\ \citenamefont {Van~Wees}}]{Cornelissen2015}%
  \BibitemOpen
  \bibfield  {author} {\bibinfo {author} {\bibfnamefont {L.}~\bibnamefont
  {Cornelissen}}, \bibinfo {author} {\bibfnamefont {J.}~\bibnamefont {Liu}},
  \bibinfo {author} {\bibfnamefont {R.}~\bibnamefont {Duine}}, \bibinfo
  {author} {\bibfnamefont {J.~B.}\ \bibnamefont {Youssef}},\ and\ \bibinfo
  {author} {\bibfnamefont {B.}~\bibnamefont {Van~Wees}},\ }\bibfield  {title}
  {\bibinfo {title} {Long-distance transport of magnon spin information in a
  magnetic insulator at room temperature},\ }\href
  {https://www.nature.com/articles/nphys3465} {\bibfield  {journal} {\bibinfo
  {journal} {Nat. Phys.}\ }\textbf {\bibinfo {volume} {11}},\ \bibinfo {pages}
  {1022} (\bibinfo {year} {2015})}\BibitemShut {NoStop}%
\bibitem [{\citenamefont {Zou}\ \emph {et~al.}(2020)\citenamefont {Zou},
  \citenamefont {Zhang},\ and\ \citenamefont {Tserkovnyak}}]{Zou2020}%
  \BibitemOpen
  \bibfield  {author} {\bibinfo {author} {\bibfnamefont {J.}~\bibnamefont
  {Zou}}, \bibinfo {author} {\bibfnamefont {S.}~\bibnamefont {Zhang}},\ and\
  \bibinfo {author} {\bibfnamefont {Y.}~\bibnamefont {Tserkovnyak}},\
  }\bibfield  {title} {\bibinfo {title} {Topological transport of deconfined
  hedgehogs in magnets},\ }\href
  {https://link.aps.org/doi/10.1103/PhysRevLett.125.267201} {\bibfield
  {journal} {\bibinfo  {journal} {Phys. Rev. Lett.}\ }\textbf {\bibinfo
  {volume} {125}},\ \bibinfo {pages} {267201} (\bibinfo {year}
  {2020})}\BibitemShut {NoStop}%
\bibitem [{\citenamefont {Mermin}(1979)}]{Mermin1979}%
  \BibitemOpen
  \bibfield  {author} {\bibinfo {author} {\bibfnamefont {N.~D.}\ \bibnamefont
  {Mermin}},\ }\bibfield  {title} {\bibinfo {title} {The topological theory of
  defects in ordered media},\ }\href
  {https://link.aps.org/doi/10.1103/RevModPhys.51.591} {\bibfield  {journal}
  {\bibinfo  {journal} {Rev. Mod. Phys.}\ }\textbf {\bibinfo {volume} {51}},\
  \bibinfo {pages} {591} (\bibinfo {year} {1979})}\BibitemShut {NoStop}%
\bibitem [{\citenamefont {de~Gennes}\ and\ \citenamefont
  {Prost}(1993)}]{Prost1993}%
  \BibitemOpen
  \bibfield  {author} {\bibinfo {author} {\bibfnamefont {P.}~\bibnamefont
  {de~Gennes}}\ and\ \bibinfo {author} {\bibfnamefont {J.}~\bibnamefont
  {Prost}},\ }\href@noop {} {\emph {\bibinfo {title} {The Physics of Liquid
  Crystals}}},\ International Series of Monographs\ (\bibinfo  {publisher}
  {Clarendon Press},\ \bibinfo {year} {1993})\BibitemShut {NoStop}%
\bibitem [{\citenamefont {Vertogen}\ and\ \citenamefont
  {Jeu}(1988)}]{Vertogen1988}%
  \BibitemOpen
  \bibfield  {author} {\bibinfo {author} {\bibfnamefont {G.}~\bibnamefont
  {Vertogen}}\ and\ \bibinfo {author} {\bibfnamefont {W.}~\bibnamefont {Jeu}},\
  }\href {https://books.google.com/books?id=Cj9CAQAAIAAJ} {\emph {\bibinfo
  {title} {Thermotropic Liquid Crystals, Fundamentals}}},\ Chemical Physics
  Series\ (\bibinfo  {publisher} {Springer-Verlag},\ \bibinfo {year}
  {1988})\BibitemShut {NoStop}%
\bibitem [{\citenamefont {Nakahara}(2003)}]{Nakahara2003}%
  \BibitemOpen
  \bibfield  {author} {\bibinfo {author} {\bibfnamefont {M.}~\bibnamefont
  {Nakahara}},\ }\href
  {http://gen.lib.rus.ec/book/index.php?md5=d8a229b3a90c803764004b69cbb457f6}
  {\emph {\bibinfo {title} {Geometry, topology, and physics}}},\ \bibinfo
  {edition} {2nd}\ ed.,\ Graduate student series in physics\ (\bibinfo
  {publisher} {Institute of Physics Publishing},\ \bibinfo {year}
  {2003})\BibitemShut {NoStop}%
\bibitem [{\citenamefont {Kleman}\ and\ \citenamefont
  {Lavrentovich}(2006)}]{Lavrentovich2006}%
  \BibitemOpen
  \bibfield  {author} {\bibinfo {author} {\bibfnamefont {M.}~\bibnamefont
  {Kleman}}\ and\ \bibinfo {author} {\bibfnamefont {O.~D.}\ \bibnamefont
  {Lavrentovich}},\ }\bibfield  {title} {\bibinfo {title} {Topological point
  defects in nematic liquid crystals},\ }\href
  {https://doi.org/10.1080/14786430600593016} {\bibfield  {journal} {\bibinfo
  {journal} {Philos. Mag.}\ }\textbf {\bibinfo {volume} {86}},\ \bibinfo
  {pages} {4117} (\bibinfo {year} {2006})}\BibitemShut {NoStop}%
\bibitem [{\citenamefont {Müller}\ and\ \citenamefont
  {Senior}(2009)}]{archscrew}%
  \BibitemOpen
  \bibfield  {author} {\bibinfo {author} {\bibfnamefont {G.}~\bibnamefont
  {Müller}}\ and\ \bibinfo {author} {\bibfnamefont {J.}~\bibnamefont
  {Senior}},\ }\bibfield  {title} {\bibinfo {title} {Simplified theory of
  {A}rchimedean screws},\ }\href {https://doi.org/10.3826/jhr.2009.3475}
  {\bibfield  {journal} {\bibinfo  {journal} {J. Hydraul. Res.}\ }\textbf
  {\bibinfo {volume} {47}},\ \bibinfo {pages} {666} (\bibinfo {year}
  {2009})}\BibitemShut {NoStop}%
\bibitem [{\citenamefont {Ichimura}\ \emph {et~al.}(1988)\citenamefont
  {Ichimura}, \citenamefont {Suzuki}, \citenamefont {Seki}, \citenamefont
  {Hosoki},\ and\ \citenamefont {Aoki}}]{Koso:1988}%
  \BibitemOpen
  \bibfield  {author} {\bibinfo {author} {\bibfnamefont {K.}~\bibnamefont
  {Ichimura}}, \bibinfo {author} {\bibfnamefont {Y.}~\bibnamefont {Suzuki}},
  \bibinfo {author} {\bibfnamefont {T.}~\bibnamefont {Seki}}, \bibinfo {author}
  {\bibfnamefont {A.}~\bibnamefont {Hosoki}},\ and\ \bibinfo {author}
  {\bibfnamefont {K.}~\bibnamefont {Aoki}},\ }\bibfield  {title} {\bibinfo
  {title} {Reversible change in alignment mode of nematic liquid crystals
  regulated photochemically by command surfaces modified with an azobenzene
  monolayer},\ }\href {https://doi.org/10.1021/la00083a030} {\bibfield
  {journal} {\bibinfo  {journal} {Langmuir}\ }\textbf {\bibinfo {volume} {4}},\
  \bibinfo {pages} {1214} (\bibinfo {year} {1988})}\BibitemShut {NoStop}%
\bibitem [{\citenamefont
  {Lavrentovich}(2020{\natexlab{a}})}]{LavrentovichRev2020}%
  \BibitemOpen
  \bibfield  {author} {\bibinfo {author} {\bibfnamefont {O.~D.}\ \bibnamefont
  {Lavrentovich}},\ }\bibfield  {title} {\bibinfo {title} {Design of nematic
  liquid crystals to control microscale dynamics},\ }\href
  {https://doi.org/10.1080/21680396.2021.1919576} {\bibfield  {journal}
  {\bibinfo  {journal} {Liq. Cryst. Rev.}\ }\textbf {\bibinfo {volume} {8}},\
  \bibinfo {pages} {59} (\bibinfo {year} {2020}{\natexlab{a}})},\ \Eprint
  {https://arxiv.org/abs/https://doi.org/10.1080/21680396.2021.1919576}
  {https://doi.org/10.1080/21680396.2021.1919576} \BibitemShut {NoStop}%
\bibitem [{\citenamefont {Tovkach}\ \emph {et~al.}(2017)\citenamefont
  {Tovkach}, \citenamefont {Conklin}, \citenamefont {Calderer}, \citenamefont
  {Golovaty}, \citenamefont {Lavrentovich}, \citenamefont {Vi\~nals},\ and\
  \citenamefont {Walkington}}]{Tovkach2017}%
  \BibitemOpen
  \bibfield  {author} {\bibinfo {author} {\bibfnamefont {O.~M.}\ \bibnamefont
  {Tovkach}}, \bibinfo {author} {\bibfnamefont {C.}~\bibnamefont {Conklin}},
  \bibinfo {author} {\bibfnamefont {M.~C.}\ \bibnamefont {Calderer}}, \bibinfo
  {author} {\bibfnamefont {D.}~\bibnamefont {Golovaty}}, \bibinfo {author}
  {\bibfnamefont {O.~D.}\ \bibnamefont {Lavrentovich}}, \bibinfo {author}
  {\bibfnamefont {J.}~\bibnamefont {Vi\~nals}},\ and\ \bibinfo {author}
  {\bibfnamefont {N.~J.}\ \bibnamefont {Walkington}},\ }\bibfield  {title}
  {\bibinfo {title} {Q-tensor model for electrokinetics in nematic liquid
  crystals},\ }\href {https://link.aps.org/doi/10.1103/PhysRevFluids.2.053302}
  {\bibfield  {journal} {\bibinfo  {journal} {Phys. Rev. Fluids}\ }\textbf
  {\bibinfo {volume} {2}},\ \bibinfo {pages} {053302} (\bibinfo {year}
  {2017})}\BibitemShut {NoStop}%
\bibitem [{\citenamefont {Tovkach}\ \emph {et~al.}(2016)\citenamefont
  {Tovkach}, \citenamefont {Calderer}, \citenamefont {Golovaty}, \citenamefont
  {Lavrentovich},\ and\ \citenamefont {Walkington}}]{Tovkach2016}%
  \BibitemOpen
  \bibfield  {author} {\bibinfo {author} {\bibfnamefont {O.~M.}\ \bibnamefont
  {Tovkach}}, \bibinfo {author} {\bibfnamefont {M.~C.}\ \bibnamefont
  {Calderer}}, \bibinfo {author} {\bibfnamefont {D.}~\bibnamefont {Golovaty}},
  \bibinfo {author} {\bibfnamefont {O.}~\bibnamefont {Lavrentovich}},\ and\
  \bibinfo {author} {\bibfnamefont {N.~J.}\ \bibnamefont {Walkington}},\
  }\bibfield  {title} {\bibinfo {title} {Electro-osmosis in nematic liquid
  crystals},\ }\href {https://link.aps.org/doi/10.1103/PhysRevE.94.012702}
  {\bibfield  {journal} {\bibinfo  {journal} {Phys. Rev. E}\ }\textbf {\bibinfo
  {volume} {94}},\ \bibinfo {pages} {012702} (\bibinfo {year}
  {2016})}\BibitemShut {NoStop}%
\bibitem [{\citenamefont {Conklin}\ \emph {et~al.}(2018)\citenamefont
  {Conklin}, \citenamefont {Tovkach}, \citenamefont {Vi\~nals}, \citenamefont
  {Calderer}, \citenamefont {Golovaty}, \citenamefont {Lavrentovich},\ and\
  \citenamefont {Walkington}}]{Conklin2018}%
  \BibitemOpen
  \bibfield  {author} {\bibinfo {author} {\bibfnamefont {C.}~\bibnamefont
  {Conklin}}, \bibinfo {author} {\bibfnamefont {O.~M.}\ \bibnamefont
  {Tovkach}}, \bibinfo {author} {\bibfnamefont {J.}~\bibnamefont {Vi\~nals}},
  \bibinfo {author} {\bibfnamefont {M.~C.}\ \bibnamefont {Calderer}}, \bibinfo
  {author} {\bibfnamefont {D.}~\bibnamefont {Golovaty}}, \bibinfo {author}
  {\bibfnamefont {O.~D.}\ \bibnamefont {Lavrentovich}},\ and\ \bibinfo {author}
  {\bibfnamefont {N.~J.}\ \bibnamefont {Walkington}},\ }\bibfield  {title}
  {\bibinfo {title} {Electrokinetic effects in nematic suspensions:
  Single-particle electro-osmosis and interparticle interactions},\ }\href
  {https://link.aps.org/doi/10.1103/PhysRevE.98.022703} {\bibfield  {journal}
  {\bibinfo  {journal} {Phys. Rev. E}\ }\textbf {\bibinfo {volume} {98}},\
  \bibinfo {pages} {022703} (\bibinfo {year} {2018})}\BibitemShut {NoStop}%
\bibitem [{\citenamefont {Lehmann}(1900)}]{lehmann1900}%
  \BibitemOpen
  \bibfield  {author} {\bibinfo {author} {\bibfnamefont {O.}~\bibnamefont
  {Lehmann}},\ }\bibfield  {title} {\bibinfo {title} {Structur, system und
  magnetisches verhalten fl{\"u}ssiger krystalle und deren mischbarkeit mit
  festen},\ }\href
  {https://onlinelibrary.wiley.com/doi/abs/10.1002/andp.19003070802} {\bibfield
   {journal} {\bibinfo  {journal} {Ann. Phys. (Berl.)}\ }\textbf {\bibinfo
  {volume} {307}},\ \bibinfo {pages} {649} (\bibinfo {year}
  {1900})}\BibitemShut {NoStop}%
\bibitem [{\citenamefont {Oswald}\ \emph {et~al.}(2019)\citenamefont {Oswald},
  \citenamefont {Dequidt},\ and\ \citenamefont {Poy}}]{Poy:2019}%
  \BibitemOpen
  \bibfield  {author} {\bibinfo {author} {\bibfnamefont {P.}~\bibnamefont
  {Oswald}}, \bibinfo {author} {\bibfnamefont {A.}~\bibnamefont {Dequidt}},\
  and\ \bibinfo {author} {\bibfnamefont {G.}~\bibnamefont {Poy}},\ }\bibfield
  {title} {\bibinfo {title} {Lehmann effect in nematic and cholesteric liquid
  crystals: a review},\ }\href
  {https://www.tandfonline.com/doi/full/10.1080/21680396.2019.1671244}
  {\bibfield  {journal} {\bibinfo  {journal} {Liq. Cryst. Rev.}\ }\textbf
  {\bibinfo {volume} {7}},\ \bibinfo {pages} {142} (\bibinfo {year}
  {2019})}\BibitemShut {NoStop}%
\bibitem [{\citenamefont {Madhusudana}\ and\ \citenamefont
  {Pratibha}(1989)}]{madhusudana:1989}%
  \BibitemOpen
  \bibfield  {author} {\bibinfo {author} {\bibfnamefont {N.}~\bibnamefont
  {Madhusudana}}\ and\ \bibinfo {author} {\bibfnamefont {R.}~\bibnamefont
  {Pratibha}},\ }\bibfield  {title} {\bibinfo {title} {An experimental
  investigation of electromechanical coupling in cholesteric liquid crystals},\
  }\href {https://doi.org/10.1080/02678298908045691} {\bibfield  {journal}
  {\bibinfo  {journal} {Liq. Cryst.}\ }\textbf {\bibinfo {volume} {5}},\
  \bibinfo {pages} {1827} (\bibinfo {year} {1989})}\BibitemShut {NoStop}%
\bibitem [{\citenamefont {Sven{\v{s}}ek}\ \emph {et~al.}(2006)\citenamefont
  {Sven{\v{s}}ek}, \citenamefont {Pleiner},\ and\ \citenamefont
  {Brand}}]{brand2006}%
  \BibitemOpen
  \bibfield  {author} {\bibinfo {author} {\bibfnamefont {D.}~\bibnamefont
  {Sven{\v{s}}ek}}, \bibinfo {author} {\bibfnamefont {H.}~\bibnamefont
  {Pleiner}},\ and\ \bibinfo {author} {\bibfnamefont {H.~R.}\ \bibnamefont
  {Brand}},\ }\bibfield  {title} {\bibinfo {title} {Phase winding in chiral
  liquid crystalline monolayers due to lehmann effects},\ }\href
  {https://doi.org/10.1103/PhysRevLett.96.140601} {\bibfield  {journal}
  {\bibinfo  {journal} {Phys. Rev. Lett.}\ }\textbf {\bibinfo {volume} {96}},\
  \bibinfo {pages} {140601} (\bibinfo {year} {2006})}\BibitemShut {NoStop}%
\bibitem [{\citenamefont {Tabe}\ and\ \citenamefont
  {Yokoyama}(2003)}]{tabe2003}%
  \BibitemOpen
  \bibfield  {author} {\bibinfo {author} {\bibfnamefont {Y.}~\bibnamefont
  {Tabe}}\ and\ \bibinfo {author} {\bibfnamefont {H.}~\bibnamefont
  {Yokoyama}},\ }\bibfield  {title} {\bibinfo {title} {Coherent collective
  precession of molecular rotors with chiral propellers},\ }\href
  {https://doi.org/10.1038/nmat1017} {\bibfield  {journal} {\bibinfo  {journal}
  {Nat. Mater.}\ }\textbf {\bibinfo {volume} {2}},\ \bibinfo {pages} {806}
  (\bibinfo {year} {2003})}\BibitemShut {NoStop}%
\bibitem [{\citenamefont {Sven\ifmmode~\check{s}\else \v{s}\fi{}ek}\ \emph
  {et~al.}(2008)\citenamefont {Sven\ifmmode~\check{s}\else \v{s}\fi{}ek},
  \citenamefont {Pleiner},\ and\ \citenamefont {Brand}}]{Svensek2008}%
  \BibitemOpen
  \bibfield  {author} {\bibinfo {author} {\bibfnamefont {D.}~\bibnamefont
  {Sven\ifmmode~\check{s}\else \v{s}\fi{}ek}}, \bibinfo {author} {\bibfnamefont
  {H.}~\bibnamefont {Pleiner}},\ and\ \bibinfo {author} {\bibfnamefont {H.~R.}\
  \bibnamefont {Brand}},\ }\bibfield  {title} {\bibinfo {title} {Inverse
  lehmann effects can be used as a microscopic pump},\ }\href
  {https://link.aps.org/doi/10.1103/PhysRevE.78.021703} {\bibfield  {journal}
  {\bibinfo  {journal} {Phys. Rev. E}\ }\textbf {\bibinfo {volume} {78}},\
  \bibinfo {pages} {021703} (\bibinfo {year} {2008})}\BibitemShut {NoStop}%
\bibitem [{sup()}]{supmat}%
  \BibitemOpen
  \href@noop {} {}\bibinfo {note} {See Supplemental Material at [] for (i) the
  generalized theory for nematic electrolytes with hydrodynamic effects
  restored, (ii) expanded discussion of the artificial symmetries, and (iii)
  proof that flexoelectric effects can be qualitatively distinguished from the
  dissipative nemato-ionic coupling.}\BibitemShut {Stop}%
\bibitem [{osm()}]{osmotic}%
  \BibitemOpen
  \href@noop {} {}\bibinfo {note} {Generically, whenever the charge transport
  is electrically triggered, the mass transport is also automatically induced.
  This, in turn, generates a corresponding dissipative field acting on the
  nematic texture. The dissipative field and motive force stemming from the
  mass sector would not qualitatively affect the kinematic effects due to the
  charge sector and could easily be included.}\BibitemShut {Stop}%
\bibitem [{\citenamefont {Zakharov}\ \emph {et~al.}(1999)\citenamefont
  {Zakharov}, \citenamefont {Komolkin},\ and\ \citenamefont
  {Maliniak}}]{Zakharov1999}%
  \BibitemOpen
  \bibfield  {author} {\bibinfo {author} {\bibfnamefont {A.~V.}\ \bibnamefont
  {Zakharov}}, \bibinfo {author} {\bibfnamefont {A.~V.}\ \bibnamefont
  {Komolkin}},\ and\ \bibinfo {author} {\bibfnamefont {A.}~\bibnamefont
  {Maliniak}},\ }\bibfield  {title} {\bibinfo {title} {Rotational viscosity in
  a nematic liquid crystal: A theoretical treatment and molecular dynamics
  simulation},\ }\href {https://link.aps.org/doi/10.1103/PhysRevE.59.6802}
  {\bibfield  {journal} {\bibinfo  {journal} {Phys. Rev. E}\ }\textbf {\bibinfo
  {volume} {59}},\ \bibinfo {pages} {6802} (\bibinfo {year}
  {1999})}\BibitemShut {NoStop}%
\bibitem [{\citenamefont {Ralph}\ and\ \citenamefont
  {Stiles}(2008)}]{Ralph2008}%
  \BibitemOpen
  \bibfield  {author} {\bibinfo {author} {\bibfnamefont {D.~C.}\ \bibnamefont
  {Ralph}}\ and\ \bibinfo {author} {\bibfnamefont {M.~D.}\ \bibnamefont
  {Stiles}},\ }\bibfield  {title} {\bibinfo {title} {Spin transfer torques},\
  }\href {https://doi.org/10.1016/j.jmmm.2007.12.019} {\bibfield  {journal}
  {\bibinfo  {journal} {J. Magn. Magn. Mater.}\ }\textbf {\bibinfo {volume}
  {320}},\ \bibinfo {pages} {1190} (\bibinfo {year} {2008})}\BibitemShut
  {NoStop}%
\bibitem [{\citenamefont {Tserkovnyak}\ \emph {et~al.}(2002)\citenamefont
  {Tserkovnyak}, \citenamefont {Brataas},\ and\ \citenamefont
  {Bauer}}]{Tserkovnyak2002}%
  \BibitemOpen
  \bibfield  {author} {\bibinfo {author} {\bibfnamefont {Y.}~\bibnamefont
  {Tserkovnyak}}, \bibinfo {author} {\bibfnamefont {A.}~\bibnamefont
  {Brataas}},\ and\ \bibinfo {author} {\bibfnamefont {G.~E.~W.}\ \bibnamefont
  {Bauer}},\ }\bibfield  {title} {\bibinfo {title} {Enhanced gilbert damping in
  thin ferromagnetic films},\ }\href
  {https://link.aps.org/doi/10.1103/PhysRevLett.88.117601} {\bibfield
  {journal} {\bibinfo  {journal} {Phys. Rev. Lett.}\ }\textbf {\bibinfo
  {volume} {88}},\ \bibinfo {pages} {117601} (\bibinfo {year}
  {2002})}\BibitemShut {NoStop}%
\bibitem [{\citenamefont {Qian}\ and\ \citenamefont {Sheng}(1998)}]{Qian1998}%
  \BibitemOpen
  \bibfield  {author} {\bibinfo {author} {\bibfnamefont {T.}~\bibnamefont
  {Qian}}\ and\ \bibinfo {author} {\bibfnamefont {P.}~\bibnamefont {Sheng}},\
  }\bibfield  {title} {\bibinfo {title} {Generalized hydrodynamic equations for
  nematic liquid crystals},\ }\href
  {https://link.aps.org/doi/10.1103/PhysRevE.58.7475} {\bibfield  {journal}
  {\bibinfo  {journal} {Phys. Rev. E}\ }\textbf {\bibinfo {volume} {58}},\
  \bibinfo {pages} {7475} (\bibinfo {year} {1998})}\BibitemShut {NoStop}%
\bibitem [{\citenamefont {Chandragiri}\ \emph {et~al.}(2020)\citenamefont
  {Chandragiri}, \citenamefont {Doostmohammadi}, \citenamefont {Yeomans},\ and\
  \citenamefont {Thampi}}]{Chandragiri2020}%
  \BibitemOpen
  \bibfield  {author} {\bibinfo {author} {\bibfnamefont {S.}~\bibnamefont
  {Chandragiri}}, \bibinfo {author} {\bibfnamefont {A.}~\bibnamefont
  {Doostmohammadi}}, \bibinfo {author} {\bibfnamefont {J.~M.}\ \bibnamefont
  {Yeomans}},\ and\ \bibinfo {author} {\bibfnamefont {S.~P.}\ \bibnamefont
  {Thampi}},\ }\bibfield  {title} {\bibinfo {title} {Flow states and
  transitions of an active nematic in a three-dimensional channel},\ }\href
  {https://link.aps.org/doi/10.1103/PhysRevLett.125.148002} {\bibfield
  {journal} {\bibinfo  {journal} {Phys. Rev. Lett.}\ }\textbf {\bibinfo
  {volume} {125}},\ \bibinfo {pages} {148002} (\bibinfo {year}
  {2020})}\BibitemShut {NoStop}%
\bibitem [{\citenamefont {Hernández-Ortiz}\ \emph {et~al.}(2011)\citenamefont
  {Hernández-Ortiz}, \citenamefont {Gettelfinger}, \citenamefont
  {Moreno-Razo},\ and\ \citenamefont {de~Pablo}}]{HernandezOrtiz2011}%
  \BibitemOpen
  \bibfield  {author} {\bibinfo {author} {\bibfnamefont {J.~P.}\ \bibnamefont
  {Hernández-Ortiz}}, \bibinfo {author} {\bibfnamefont {B.~T.}\ \bibnamefont
  {Gettelfinger}}, \bibinfo {author} {\bibfnamefont {J.}~\bibnamefont
  {Moreno-Razo}},\ and\ \bibinfo {author} {\bibfnamefont {J.~J.}\ \bibnamefont
  {de~Pablo}},\ }\bibfield  {title} {\bibinfo {title} {Modeling flows of
  confined nematic liquid crystals},\ }\href
  {https://doi.org/10.1063/1.3567098} {\bibfield  {journal} {\bibinfo
  {journal} {J. Chem. Phys.}\ }\textbf {\bibinfo {volume} {134}},\ \bibinfo
  {pages} {134905} (\bibinfo {year} {2011})}\BibitemShut {NoStop}%
\bibitem [{\citenamefont {Leslie}(1979)}]{Leslie1979}%
  \BibitemOpen
  \bibfield  {author} {\bibinfo {author} {\bibfnamefont {F.}~\bibnamefont
  {Leslie}},\ }\bibfield  {title} {\bibinfo {title} {Theory of flow phenomena
  in liquid crystals}\ }(\bibinfo  {publisher} {Elsevier},\ \bibinfo {year}
  {1979})\ pp.\ \bibinfo {pages} {1--81}\BibitemShut {NoStop}%
\bibitem [{\citenamefont {Calderer}\ \emph {et~al.}(2016)\citenamefont
  {Calderer}, \citenamefont {Golovaty}, \citenamefont {Lavrentovich},\ and\
  \citenamefont {Walkington}}]{Calderer2016}%
  \BibitemOpen
  \bibfield  {author} {\bibinfo {author} {\bibfnamefont {M.~C.}\ \bibnamefont
  {Calderer}}, \bibinfo {author} {\bibfnamefont {D.}~\bibnamefont {Golovaty}},
  \bibinfo {author} {\bibfnamefont {O.}~\bibnamefont {Lavrentovich}},\ and\
  \bibinfo {author} {\bibfnamefont {N.~J.}\ \bibnamefont {Walkington}},\
  }\bibfield  {title} {\bibinfo {title} {Modeling of nematic electrolytes and
  nonlinear electroosmosis},\ }\href {http://www.jstor.org/stable/26166774}
  {\bibfield  {journal} {\bibinfo  {journal} {SIAM J. Appl. Math.}\ }\textbf
  {\bibinfo {volume} {76}},\ \bibinfo {pages} {2260} (\bibinfo {year}
  {2016})}\BibitemShut {NoStop}%
\bibitem [{\citenamefont {Ball}(2017)}]{doi:10.1080/15421406.2017.1289425}%
  \BibitemOpen
  \bibfield  {author} {\bibinfo {author} {\bibfnamefont {J.~M.}\ \bibnamefont
  {Ball}},\ }\bibfield  {title} {\bibinfo {title} {Mathematics and liquid
  crystals},\ }\href {https://doi.org/10.1080/15421406.2017.1289425} {\bibfield
   {journal} {\bibinfo  {journal} {Mol. Cryst. Liq. Cryst.}\ }\textbf {\bibinfo
  {volume} {647}},\ \bibinfo {pages} {1} (\bibinfo {year} {2017})}\BibitemShut
  {NoStop}%
\bibitem [{\citenamefont {Chandrasekhar}(1992)}]{chandrasekhar_1992}%
  \BibitemOpen
  \bibfield  {author} {\bibinfo {author} {\bibfnamefont {S.}~\bibnamefont
  {Chandrasekhar}},\ }\href {https://doi.org/10.1017/CBO9780511622496} {\emph
  {\bibinfo {title} {Liquid Crystals}}},\ \bibinfo {edition} {2nd}\ ed.\
  (\bibinfo  {publisher} {Cambridge University Press},\ \bibinfo {year}
  {1992})\BibitemShut {NoStop}%
\bibitem [{\citenamefont {Shen}\ and\ \citenamefont {Hu}(2010)}]{Shen2010}%
  \BibitemOpen
  \bibfield  {author} {\bibinfo {author} {\bibfnamefont {S.}~\bibnamefont
  {Shen}}\ and\ \bibinfo {author} {\bibfnamefont {S.}~\bibnamefont {Hu}},\
  }\bibfield  {title} {\bibinfo {title} {A theory of flexoelectricity with
  surface effect for elastic dielectrics},\ }\href
  {https://www.sciencedirect.com/science/article/pii/S0022509610000499}
  {\bibfield  {journal} {\bibinfo  {journal} {J. Mech. Phys. Solids}\ }\textbf
  {\bibinfo {volume} {58}},\ \bibinfo {pages} {665} (\bibinfo {year}
  {2010})}\BibitemShut {NoStop}%
\bibitem [{\citenamefont {Gyarmati}\ \emph {et~al.}(1970)\citenamefont
  {Gyarmati} \emph {et~al.}}]{Gyarmati1970}%
  \BibitemOpen
  \bibfield  {author} {\bibinfo {author} {\bibfnamefont {I.}~\bibnamefont
  {Gyarmati}} \emph {et~al.},\ }\href@noop {} {\emph {\bibinfo {title}
  {Non-equilibrium thermodynamics}}},\ Vol.\ \bibinfo {volume} {184}\ (\bibinfo
   {publisher} {Springer},\ \bibinfo {year} {1970})\BibitemShut {NoStop}%
\bibitem [{\citenamefont {Sonnet}\ \emph {et~al.}(2004)\citenamefont {Sonnet},
  \citenamefont {Maffettone},\ and\ \citenamefont {Virga}}]{Sonnet:2004}%
  \BibitemOpen
  \bibfield  {author} {\bibinfo {author} {\bibfnamefont {A.}~\bibnamefont
  {Sonnet}}, \bibinfo {author} {\bibfnamefont {P.}~\bibnamefont {Maffettone}},\
  and\ \bibinfo {author} {\bibfnamefont {E.}~\bibnamefont {Virga}},\ }\bibfield
   {title} {\bibinfo {title} {Continuum theory for nematic liquid crystals with
  tensorial order},\ }\href
  {https://www.sciencedirect.com/science/article/pii/S0377025704000886}
  {\bibfield  {journal} {\bibinfo  {journal} {J. Non-Newton. Fluid Mech.}\
  }\textbf {\bibinfo {volume} {119}},\ \bibinfo {pages} {51} (\bibinfo {year}
  {2004})}\BibitemShut {NoStop}%
\bibitem [{\citenamefont {Onsager}(1931)}]{Onsager1931}%
  \BibitemOpen
  \bibfield  {author} {\bibinfo {author} {\bibfnamefont {L.}~\bibnamefont
  {Onsager}},\ }\bibfield  {title} {\bibinfo {title} {Reciprocal relations in
  irreversible processes. i.},\ }\href
  {https://link.aps.org/doi/10.1103/PhysRev.37.405} {\bibfield  {journal}
  {\bibinfo  {journal} {Phys. Rev.}\ }\textbf {\bibinfo {volume} {37}},\
  \bibinfo {pages} {405} (\bibinfo {year} {1931})}\BibitemShut {NoStop}%
\bibitem [{\citenamefont {Matsumoto}\ \emph {et~al.}(1976)\citenamefont
  {Matsumoto}, \citenamefont {Kawamoto},\ and\ \citenamefont
  {Mizunoya}}]{Matsumoto1976}%
  \BibitemOpen
  \bibfield  {author} {\bibinfo {author} {\bibfnamefont {S.}~\bibnamefont
  {Matsumoto}}, \bibinfo {author} {\bibfnamefont {M.}~\bibnamefont
  {Kawamoto}},\ and\ \bibinfo {author} {\bibfnamefont {K.}~\bibnamefont
  {Mizunoya}},\ }\bibfield  {title} {\bibinfo {title} {Field‐induced
  deformation of hybrid‐aligned nematic liquid crystals: New multicolor
  liquid crystal display},\ }\href {https://doi.org/10.1063/1.323245}
  {\bibfield  {journal} {\bibinfo  {journal} {J. Appl. Phys.}\ }\textbf
  {\bibinfo {volume} {47}},\ \bibinfo {pages} {3842} (\bibinfo {year}
  {1976})}\BibitemShut {NoStop}%
\bibitem [{\citenamefont {Frisken}\ and\ \citenamefont
  {Palffy-Muhoray}(1989)}]{Frisken:1989}%
  \BibitemOpen
  \bibfield  {author} {\bibinfo {author} {\bibfnamefont {B.~J.}\ \bibnamefont
  {Frisken}}\ and\ \bibinfo {author} {\bibfnamefont {P.}~\bibnamefont
  {Palffy-Muhoray}},\ }\bibfield  {title} {\bibinfo {title} {Freedericksz
  transitions in nematic liquid crystals: The effects of an in-plane electric
  field},\ }\href {https://doi.org/10.1103/PhysRevA.40.6099} {\bibfield
  {journal} {\bibinfo  {journal} {Phys. Rev. A}\ }\textbf {\bibinfo {volume}
  {40}},\ \bibinfo {pages} {6099} (\bibinfo {year} {1989})}\BibitemShut
  {NoStop}%
\bibitem [{imp()}]{impedance}%
  \BibitemOpen
  \href@noop {} {}\bibinfo {note} {When the frequency of the AC field is
  comparable to the natural frequency associated with nematic dynamics, the
  induced impedance will have capacitative contributions in addition to
  inductive contributions. Capacitative contributions also arise when screening
  due to charge buildup at the boundaries is considered.}\BibitemShut {Stop}%
\bibitem [{\citenamefont {Sandford~O’Neill}\ \emph
  {et~al.}(2020)\citenamefont {Sandford~O’Neill}, \citenamefont {Salter},
  \citenamefont {Booth}, \citenamefont {Elston},\ and\ \citenamefont
  {Morris}}]{Sandford2020}%
  \BibitemOpen
  \bibfield  {author} {\bibinfo {author} {\bibfnamefont {J.~J.}\ \bibnamefont
  {Sandford~O’Neill}}, \bibinfo {author} {\bibfnamefont {P.~S.}\ \bibnamefont
  {Salter}}, \bibinfo {author} {\bibfnamefont {M.~J.}\ \bibnamefont {Booth}},
  \bibinfo {author} {\bibfnamefont {S.~J.}\ \bibnamefont {Elston}},\ and\
  \bibinfo {author} {\bibfnamefont {S.~M.}\ \bibnamefont {Morris}},\ }\bibfield
   {title} {\bibinfo {title} {Electrically-tunable positioning of topological
  defects in liquid crystals},\ }\href
  {https://doi.org/10.1038/s41467-020-16059-1} {\bibfield  {journal} {\bibinfo
  {journal} {Nat. Commun.}\ }\textbf {\bibinfo {volume} {11}},\ \bibinfo
  {pages} {1} (\bibinfo {year} {2020})}\BibitemShut {NoStop}%
\bibitem [{\citenamefont {Thiele}(1973)}]{Thiele1973}%
  \BibitemOpen
  \bibfield  {author} {\bibinfo {author} {\bibfnamefont {A.~A.}\ \bibnamefont
  {Thiele}},\ }\bibfield  {title} {\bibinfo {title} {Steady-state motion of
  magnetic domains},\ }\href
  {https://link.aps.org/doi/10.1103/PhysRevLett.30.230} {\bibfield  {journal}
  {\bibinfo  {journal} {Phys. Rev. Lett.}\ }\textbf {\bibinfo {volume} {30}},\
  \bibinfo {pages} {230} (\bibinfo {year} {1973})}\BibitemShut {NoStop}%
\bibitem [{\citenamefont {Imura}\ and\ \citenamefont
  {Okano}(1973)}]{Imura1973}%
  \BibitemOpen
  \bibfield  {author} {\bibinfo {author} {\bibfnamefont {H.}~\bibnamefont
  {Imura}}\ and\ \bibinfo {author} {\bibfnamefont {K.}~\bibnamefont {Okano}},\
  }\bibfield  {title} {\bibinfo {title} {Friction coefficient for a moving
  disinclination in a nematic liquid crystal},\ }\href
  {https://www.sciencedirect.com/science/article/pii/0375960173907287}
  {\bibfield  {journal} {\bibinfo  {journal} {Phys. Lett. A}\ }\textbf
  {\bibinfo {volume} {42}},\ \bibinfo {pages} {403} (\bibinfo {year}
  {1973})}\BibitemShut {NoStop}%
\bibitem [{\citenamefont {Tang}\ and\ \citenamefont
  {Selinger}(2019)}]{Tang2019}%
  \BibitemOpen
  \bibfield  {author} {\bibinfo {author} {\bibfnamefont {X.}~\bibnamefont
  {Tang}}\ and\ \bibinfo {author} {\bibfnamefont {J.~V.}\ \bibnamefont
  {Selinger}},\ }\bibfield  {title} {\bibinfo {title} {Theory of defect motion
  in 2d passive and active nematic liquid crystals},\ }\href
  {https://doi.org/10.1039/C8SM01901K} {\bibfield  {journal} {\bibinfo
  {journal} {Soft Matter}\ }\textbf {\bibinfo {volume} {15}},\ \bibinfo {pages}
  {587} (\bibinfo {year} {2019})}\BibitemShut {NoStop}%
\bibitem [{\citenamefont {Yoshida}\ \emph {et~al.}(2015)\citenamefont
  {Yoshida}, \citenamefont {Asakura}, \citenamefont {Fukuda},\ and\
  \citenamefont {Ozaki}}]{Yoshida2015}%
  \BibitemOpen
  \bibfield  {author} {\bibinfo {author} {\bibfnamefont {H.}~\bibnamefont
  {Yoshida}}, \bibinfo {author} {\bibfnamefont {K.}~\bibnamefont {Asakura}},
  \bibinfo {author} {\bibfnamefont {J.-i.}\ \bibnamefont {Fukuda}},\ and\
  \bibinfo {author} {\bibfnamefont {M.}~\bibnamefont {Ozaki}},\ }\bibfield
  {title} {\bibinfo {title} {Three-dimensional positioning and control of
  colloidal objects utilizing engineered liquid crystalline defect networks},\
  }\href {https://doi.org/10.1038/ncomms8180} {\bibfield  {journal} {\bibinfo
  {journal} {Nat. Commun.}\ }\textbf {\bibinfo {volume} {6}},\ \bibinfo {pages}
  {1} (\bibinfo {year} {2015})}\BibitemShut {NoStop}%
\bibitem [{\citenamefont {Ghadimi~Nassiri}\ and\ \citenamefont
  {Brasselet}(2018)}]{Nassiri2018}%
  \BibitemOpen
  \bibfield  {author} {\bibinfo {author} {\bibfnamefont {M.}~\bibnamefont
  {Ghadimi~Nassiri}}\ and\ \bibinfo {author} {\bibfnamefont {E.}~\bibnamefont
  {Brasselet}},\ }\bibfield  {title} {\bibinfo {title} {Multispectral
  management of the photon orbital angular momentum},\ }\href
  {https://link.aps.org/doi/10.1103/PhysRevLett.121.213901} {\bibfield
  {journal} {\bibinfo  {journal} {Phys. Rev. Lett.}\ }\textbf {\bibinfo
  {volume} {121}},\ \bibinfo {pages} {213901} (\bibinfo {year}
  {2018})}\BibitemShut {NoStop}%
\bibitem [{\citenamefont {Patil}\ \emph {et~al.}(2020)\citenamefont {Patil},
  \citenamefont {Kos}, \citenamefont {Ravnik},\ and\ \citenamefont
  {Dunkel}}]{Patil2020}%
  \BibitemOpen
  \bibfield  {author} {\bibinfo {author} {\bibfnamefont {V.~P.}\ \bibnamefont
  {Patil}}, \bibinfo {author} {\bibfnamefont {{\v{Z}}.}~\bibnamefont {Kos}},
  \bibinfo {author} {\bibfnamefont {M.}~\bibnamefont {Ravnik}},\ and\ \bibinfo
  {author} {\bibfnamefont {J.}~\bibnamefont {Dunkel}},\ }\bibfield  {title}
  {\bibinfo {title} {Discharging dynamics of topological batteries},\ }\href
  {https://link.aps.org/doi/10.1103/PhysRevResearch.2.043196} {\bibfield
  {journal} {\bibinfo  {journal} {Phys. Rev. Res.}\ }\textbf {\bibinfo {volume}
  {2}},\ \bibinfo {pages} {043196} (\bibinfo {year} {2020})}\BibitemShut
  {NoStop}%
\bibitem [{\citenamefont {Tserkovnyak}\ and\ \citenamefont
  {Xiao}(2018)}]{Tserkovnyak2018}%
  \BibitemOpen
  \bibfield  {author} {\bibinfo {author} {\bibfnamefont {Y.}~\bibnamefont
  {Tserkovnyak}}\ and\ \bibinfo {author} {\bibfnamefont {J.}~\bibnamefont
  {Xiao}},\ }\bibfield  {title} {\bibinfo {title} {Energy storage via
  topological spin textures},\ }\href
  {https://doi.org/10.1103/PhysRevLett.121.127701} {\bibfield  {journal}
  {\bibinfo  {journal} {Phys. Rev. Lett.}\ }\textbf {\bibinfo {volume} {121}},\
  \bibinfo {pages} {127701} (\bibinfo {year} {2018})}\BibitemShut {NoStop}%
\bibitem [{\citenamefont {Žiga Kos}\ and\ \citenamefont
  {Dunkel}(2022)}]{Kos2022}%
  \BibitemOpen
  \bibfield  {author} {\bibinfo {author} {\bibnamefont {Žiga Kos}}\ and\
  \bibinfo {author} {\bibfnamefont {J.}~\bibnamefont {Dunkel}},\ }\bibfield
  {title} {\bibinfo {title} {Nematic bits and universal logic gates},\ }\href
  {https://www.science.org/doi/abs/10.1126/sciadv.abp8371} {\bibfield
  {journal} {\bibinfo  {journal} {Sci. Adv.}\ }\textbf {\bibinfo {volume}
  {8}},\ \bibinfo {pages} {eabp8371} (\bibinfo {year} {2022})}\BibitemShut
  {NoStop}%
\bibitem [{\citenamefont {J\'akli}\ \emph {et~al.}(2018)\citenamefont
  {J\'akli}, \citenamefont {Lavrentovich},\ and\ \citenamefont
  {Selinger}}]{Selinger:2018}%
  \BibitemOpen
  \bibfield  {author} {\bibinfo {author} {\bibfnamefont {A.}~\bibnamefont
  {J\'akli}}, \bibinfo {author} {\bibfnamefont {O.~D.}\ \bibnamefont
  {Lavrentovich}},\ and\ \bibinfo {author} {\bibfnamefont {J.~V.}\ \bibnamefont
  {Selinger}},\ }\bibfield  {title} {\bibinfo {title} {Physics of liquid
  crystals of bent-shaped molecules},\ }\href
  {https://doi.org/10.1103/RevModPhys.90.045004} {\bibfield  {journal}
  {\bibinfo  {journal} {Rev. Mod. Phys.}\ }\textbf {\bibinfo {volume} {90}},\
  \bibinfo {pages} {045004} (\bibinfo {year} {2018})}\BibitemShut {NoStop}%
\bibitem [{\citenamefont {Alexander}\ and\ \citenamefont
  {Yeomans}(2007)}]{Yeomans:2007}%
  \BibitemOpen
  \bibfield  {author} {\bibinfo {author} {\bibfnamefont {G.~P.}\ \bibnamefont
  {Alexander}}\ and\ \bibinfo {author} {\bibfnamefont {J.~M.}\ \bibnamefont
  {Yeomans}},\ }\bibfield  {title} {\bibinfo {title} {Flexoelectric blue
  phases},\ }\href {https://doi.org/10.1103/PhysRevLett.99.067801} {\bibfield
  {journal} {\bibinfo  {journal} {Phys. Rev. Lett.}\ }\textbf {\bibinfo
  {volume} {99}},\ \bibinfo {pages} {067801} (\bibinfo {year}
  {2007})}\BibitemShut {NoStop}%
\bibitem [{\citenamefont {Castles}\ \emph {et~al.}(2010)\citenamefont
  {Castles}, \citenamefont {Morris}, \citenamefont {Terentjev},\ and\
  \citenamefont {Coles}}]{Castles2010}%
  \BibitemOpen
  \bibfield  {author} {\bibinfo {author} {\bibfnamefont {F.}~\bibnamefont
  {Castles}}, \bibinfo {author} {\bibfnamefont {S.~M.}\ \bibnamefont {Morris}},
  \bibinfo {author} {\bibfnamefont {E.~M.}\ \bibnamefont {Terentjev}},\ and\
  \bibinfo {author} {\bibfnamefont {H.~J.}\ \bibnamefont {Coles}},\ }\bibfield
  {title} {\bibinfo {title} {Thermodynamically stable blue phases},\ }\href
  {https://link.aps.org/doi/10.1103/PhysRevLett.104.157801} {\bibfield
  {journal} {\bibinfo  {journal} {Phys. Rev. Lett.}\ }\textbf {\bibinfo
  {volume} {104}},\ \bibinfo {pages} {157801} (\bibinfo {year}
  {2010})}\BibitemShut {NoStop}%
\bibitem [{\citenamefont {Berger}(1978)}]{Berger1978}%
  \BibitemOpen
  \bibfield  {author} {\bibinfo {author} {\bibfnamefont {L.}~\bibnamefont
  {Berger}},\ }\bibfield  {title} {\bibinfo {title} {Low-field
  magnetoresistance and domain drag in ferromagnets},\ }\href
  {https://doi.org/10.1063/1.324716} {\bibfield  {journal} {\bibinfo  {journal}
  {J. Appl. Phys.}\ }\textbf {\bibinfo {volume} {49}},\ \bibinfo {pages} {2156}
  (\bibinfo {year} {1978})}\BibitemShut {NoStop}%
\bibitem [{\citenamefont {Kittel}(1949)}]{Kittel1949}%
  \BibitemOpen
  \bibfield  {author} {\bibinfo {author} {\bibfnamefont {C.}~\bibnamefont
  {Kittel}},\ }\bibfield  {title} {\bibinfo {title} {Physical theory of
  ferromagnetic domains},\ }\href
  {https://link.aps.org/doi/10.1103/RevModPhys.21.541} {\bibfield  {journal}
  {\bibinfo  {journal} {Rev. Mod. Phys.}\ }\textbf {\bibinfo {volume} {21}},\
  \bibinfo {pages} {541} (\bibinfo {year} {1949})}\BibitemShut {NoStop}%
\bibitem [{\citenamefont {Sebasti\'an}\ \emph {et~al.}(2020)\citenamefont
  {Sebasti\'an}, \citenamefont {Cmok}, \citenamefont {Mandle}, \citenamefont
  {de~la Fuente}, \citenamefont {Dreven\ifmmode \check{s}\else~\v{s}\fi{}ek
  Olenik}, \citenamefont {\ifmmode \check{C}\else
  \v{C}\fi{}opi\ifmmode~\check{c}\else \v{c}\fi{}},\ and\ \citenamefont
  {Mertelj}}]{Mertelj2020}%
  \BibitemOpen
  \bibfield  {author} {\bibinfo {author} {\bibfnamefont {N.}~\bibnamefont
  {Sebasti\'an}}, \bibinfo {author} {\bibfnamefont {L.}~\bibnamefont {Cmok}},
  \bibinfo {author} {\bibfnamefont {R.~J.}\ \bibnamefont {Mandle}}, \bibinfo
  {author} {\bibfnamefont {M.~R.}\ \bibnamefont {de~la Fuente}}, \bibinfo
  {author} {\bibfnamefont {I.}~\bibnamefont {Dreven\ifmmode
  \check{s}\else~\v{s}\fi{}ek Olenik}}, \bibinfo {author} {\bibfnamefont
  {M.}~\bibnamefont {\ifmmode \check{C}\else
  \v{C}\fi{}opi\ifmmode~\check{c}\else \v{c}\fi{}}},\ and\ \bibinfo {author}
  {\bibfnamefont {A.}~\bibnamefont {Mertelj}},\ }\bibfield  {title} {\bibinfo
  {title} {Ferroelectric-ferroelastic phase transition in a nematic liquid
  crystal},\ }\href {https://doi.org/10.1103/PhysRevLett.124.037801} {\bibfield
   {journal} {\bibinfo  {journal} {Phys. Rev. Lett.}\ }\textbf {\bibinfo
  {volume} {124}},\ \bibinfo {pages} {037801} (\bibinfo {year}
  {2020})}\BibitemShut {NoStop}%
\bibitem [{\citenamefont
  {Lavrentovich}(2020{\natexlab{b}})}]{Lavrentovich2020}%
  \BibitemOpen
  \bibfield  {author} {\bibinfo {author} {\bibfnamefont {O.~D.}\ \bibnamefont
  {Lavrentovich}},\ }\bibfield  {title} {\bibinfo {title} {Ferroelectric
  nematic liquid crystal, a century in waiting},\ }\href
  {https://doi.org/10.1073/pnas.2008947117} {\bibfield  {journal} {\bibinfo
  {journal} {Proc. Natl. Acad. Sci. U.S.A.}\ }\textbf {\bibinfo {volume}
  {117}},\ \bibinfo {pages} {14629} (\bibinfo {year}
  {2020}{\natexlab{b}})}\BibitemShut {NoStop}%
\bibitem [{\citenamefont {Warren}(2020)}]{Warren:2020}%
  \BibitemOpen
  \bibfield  {author} {\bibinfo {author} {\bibfnamefont {P.~B.}\ \bibnamefont
  {Warren}},\ }\bibfield  {title} {\bibinfo {title} {Non-{F}aradaic electric
  currents in the nernst-planck equations and nonlocal diffusiophoresis of
  suspended colloids in crossed salt gradients},\ }\href
  {https://doi.org/10.1103/PhysRevLett.124.248004} {\bibfield  {journal}
  {\bibinfo  {journal} {Phys. Rev. Lett.}\ }\textbf {\bibinfo {volume} {124}},\
  \bibinfo {pages} {248004} (\bibinfo {year} {2020})}\BibitemShut {NoStop}%
\bibitem [{\citenamefont {Avni}\ \emph {et~al.}(2019)\citenamefont {Avni},
  \citenamefont {Andelman},\ and\ \citenamefont {Podgornik}}]{Avni:2019}%
  \BibitemOpen
  \bibfield  {author} {\bibinfo {author} {\bibfnamefont {Y.}~\bibnamefont
  {Avni}}, \bibinfo {author} {\bibfnamefont {D.}~\bibnamefont {Andelman}},\
  and\ \bibinfo {author} {\bibfnamefont {R.}~\bibnamefont {Podgornik}},\
  }\bibfield  {title} {\bibinfo {title} {Charge regulation with fixed and
  mobile charged macromolecules},\ }\href
  {https://doi.org/https://doi.org/10.1016/j.coelec.2018.10.014} {\bibfield
  {journal} {\bibinfo  {journal} {Curr. Opin. Electrochem.}\ }\textbf {\bibinfo
  {volume} {13}},\ \bibinfo {pages} {70} (\bibinfo {year} {2019})}\BibitemShut
  {NoStop}%
\end{thebibliography}%


\begin{thebibliography}{11}%
\makeatletter
\providecommand \@ifxundefined [1]{%
 \@ifx{#1\undefined}
}%
\providecommand \@ifnum [1]{%
 \ifnum #1\expandafter \@firstoftwo
 \else \expandafter \@secondoftwo
 \fi
}%
\providecommand \@ifx [1]{%
 \ifx #1\expandafter \@firstoftwo
 \else \expandafter \@secondoftwo
 \fi
}%
\providecommand \natexlab [1]{#1}%
\providecommand \enquote  [1]{``#1''}%
\providecommand \bibnamefont  [1]{#1}%
\providecommand \bibfnamefont [1]{#1}%
\providecommand \citenamefont [1]{#1}%
\providecommand \href@noop [0]{\@secondoftwo}%
\providecommand \href [0]{\begingroup \@sanitize@url \@href}%
\providecommand \@href[1]{\@@startlink{#1}\@@href}%
\providecommand \@@href[1]{\endgroup#1\@@endlink}%
\providecommand \@sanitize@url [0]{\catcode `\\12\catcode `\$12\catcode
  `\&12\catcode `\#12\catcode `\^12\catcode `\_12\catcode `\%12\relax}%
\providecommand \@@startlink[1]{}%
\providecommand \@@endlink[0]{}%
\providecommand \url  [0]{\begingroup\@sanitize@url \@url }%
\providecommand \@url [1]{\endgroup\@href {#1}{\urlprefix }}%
\providecommand \urlprefix  [0]{URL }%
\providecommand \Eprint [0]{\href }%
\providecommand \doibase [0]{https://doi.org/}%
\providecommand \selectlanguage [0]{\@gobble}%
\providecommand \bibinfo  [0]{\@secondoftwo}%
\providecommand \bibfield  [0]{\@secondoftwo}%
\providecommand \translation [1]{[#1]}%
\providecommand \BibitemOpen [0]{}%
\providecommand \bibitemStop [0]{}%
\providecommand \bibitemNoStop [0]{.\EOS\space}%
\providecommand \EOS [0]{\spacefactor3000\relax}%
\providecommand \BibitemShut  [1]{\csname bibitem#1\endcsname}%
\let\auto@bib@innerbib\@empty
\bibitem [{\citenamefont {Leslie}(1979)}]{Leslie1979}%
  \BibitemOpen
  \bibfield  {author} {\bibinfo {author} {\bibfnamefont {F.}~\bibnamefont
  {Leslie}},\ }\bibfield  {title} {\bibinfo {title} {Theory of flow phenomena
  in liquid crystals}\ }(\bibinfo  {publisher} {Elsevier},\ \bibinfo {year}
  {1979})\ pp.\ \bibinfo {pages} {1--81}\BibitemShut {NoStop}%
\bibitem [{\citenamefont {Tovkach}\ \emph {et~al.}(2016)\citenamefont
  {Tovkach}, \citenamefont {Calderer}, \citenamefont {Golovaty}, \citenamefont
  {Lavrentovich},\ and\ \citenamefont {Walkington}}]{Tovkach2016}%
  \BibitemOpen
  \bibfield  {author} {\bibinfo {author} {\bibfnamefont {O.~M.}\ \bibnamefont
  {Tovkach}}, \bibinfo {author} {\bibfnamefont {M.~C.}\ \bibnamefont
  {Calderer}}, \bibinfo {author} {\bibfnamefont {D.}~\bibnamefont {Golovaty}},
  \bibinfo {author} {\bibfnamefont {O.}~\bibnamefont {Lavrentovich}},\ and\
  \bibinfo {author} {\bibfnamefont {N.~J.}\ \bibnamefont {Walkington}},\
  }\bibfield  {title} {\bibinfo {title} {Electro-osmosis in nematic liquid
  crystals},\ }\href {https://link.aps.org/doi/10.1103/PhysRevE.94.012702}
  {\bibfield  {journal} {\bibinfo  {journal} {Phys. Rev. E}\ }\textbf {\bibinfo
  {volume} {94}},\ \bibinfo {pages} {012702} (\bibinfo {year}
  {2016})}\BibitemShut {NoStop}%
\bibitem [{\citenamefont {Tovkach}\ \emph {et~al.}(2017)\citenamefont
  {Tovkach}, \citenamefont {Conklin}, \citenamefont {Calderer}, \citenamefont
  {Golovaty}, \citenamefont {Lavrentovich}, \citenamefont {Vi\~nals},\ and\
  \citenamefont {Walkington}}]{Tovkach2017}%
  \BibitemOpen
  \bibfield  {author} {\bibinfo {author} {\bibfnamefont {O.~M.}\ \bibnamefont
  {Tovkach}}, \bibinfo {author} {\bibfnamefont {C.}~\bibnamefont {Conklin}},
  \bibinfo {author} {\bibfnamefont {M.~C.}\ \bibnamefont {Calderer}}, \bibinfo
  {author} {\bibfnamefont {D.}~\bibnamefont {Golovaty}}, \bibinfo {author}
  {\bibfnamefont {O.~D.}\ \bibnamefont {Lavrentovich}}, \bibinfo {author}
  {\bibfnamefont {J.}~\bibnamefont {Vi\~nals}},\ and\ \bibinfo {author}
  {\bibfnamefont {N.~J.}\ \bibnamefont {Walkington}},\ }\bibfield  {title}
  {\bibinfo {title} {Q-tensor model for electrokinetics in nematic liquid
  crystals},\ }\href {https://link.aps.org/doi/10.1103/PhysRevFluids.2.053302}
  {\bibfield  {journal} {\bibinfo  {journal} {Phys. Rev. Fluids}\ }\textbf
  {\bibinfo {volume} {2}},\ \bibinfo {pages} {053302} (\bibinfo {year}
  {2017})}\BibitemShut {NoStop}%
\bibitem [{\citenamefont {Calderer}\ \emph {et~al.}(2016)\citenamefont
  {Calderer}, \citenamefont {Golovaty}, \citenamefont {Lavrentovich},\ and\
  \citenamefont {Walkington}}]{Calderer2016}%
  \BibitemOpen
  \bibfield  {author} {\bibinfo {author} {\bibfnamefont {M.~C.}\ \bibnamefont
  {Calderer}}, \bibinfo {author} {\bibfnamefont {D.}~\bibnamefont {Golovaty}},
  \bibinfo {author} {\bibfnamefont {O.}~\bibnamefont {Lavrentovich}},\ and\
  \bibinfo {author} {\bibfnamefont {N.~J.}\ \bibnamefont {Walkington}},\
  }\bibfield  {title} {\bibinfo {title} {Modeling of nematic electrolytes and
  nonlinear electroosmosis},\ }\href {http://www.jstor.org/stable/26166774}
  {\bibfield  {journal} {\bibinfo  {journal} {SIAM J. Appl. Math.}\ }\textbf
  {\bibinfo {volume} {76}},\ \bibinfo {pages} {2260} (\bibinfo {year}
  {2016})}\BibitemShut {NoStop}%
\bibitem [{\citenamefont {Gyarmati}\ \emph {et~al.}(1970)\citenamefont
  {Gyarmati} \emph {et~al.}}]{Gyarmati1970}%
  \BibitemOpen
  \bibfield  {author} {\bibinfo {author} {\bibfnamefont {I.}~\bibnamefont
  {Gyarmati}} \emph {et~al.},\ }\href@noop {} {\emph {\bibinfo {title}
  {Non-equilibrium thermodynamics}}},\ Vol.\ \bibinfo {volume} {184}\ (\bibinfo
   {publisher} {Springer},\ \bibinfo {year} {1970})\BibitemShut {NoStop}%
\bibitem [{\citenamefont {Chandrasekhar}(1992)}]{chandrasekhar_1992}%
  \BibitemOpen
  \bibfield  {author} {\bibinfo {author} {\bibfnamefont {S.}~\bibnamefont
  {Chandrasekhar}},\ }\href {https://doi.org/10.1017/CBO9780511622496} {\emph
  {\bibinfo {title} {Liquid Crystals}}},\ \bibinfo {edition} {2nd}\ ed.\
  (\bibinfo  {publisher} {Cambridge University Press},\ \bibinfo {year}
  {1992})\BibitemShut {NoStop}%
\bibitem [{\citenamefont {de~Gennes}\ and\ \citenamefont
  {Prost}(1993)}]{Prost1993}%
  \BibitemOpen
  \bibfield  {author} {\bibinfo {author} {\bibfnamefont {P.}~\bibnamefont
  {de~Gennes}}\ and\ \bibinfo {author} {\bibfnamefont {J.}~\bibnamefont
  {Prost}},\ }\href@noop {} {\emph {\bibinfo {title} {The Physics of Liquid
  Crystals}}},\ International Series of Monographs\ (\bibinfo  {publisher}
  {Clarendon Press},\ \bibinfo {year} {1993})\BibitemShut {NoStop}%
\bibitem [{\citenamefont {Vertogen}\ and\ \citenamefont
  {Jeu}(1988)}]{Vertogen1988}%
  \BibitemOpen
  \bibfield  {author} {\bibinfo {author} {\bibfnamefont {G.}~\bibnamefont
  {Vertogen}}\ and\ \bibinfo {author} {\bibfnamefont {W.}~\bibnamefont {Jeu}},\
  }\href {https://books.google.com/books?id=Cj9CAQAAIAAJ} {\emph {\bibinfo
  {title} {Thermotropic Liquid Crystals, Fundamentals}}},\ Chemical Physics
  Series\ (\bibinfo  {publisher} {Springer-Verlag},\ \bibinfo {year}
  {1988})\BibitemShut {NoStop}%
\bibitem [{\citenamefont {Zakharov}\ \emph {et~al.}(1999)\citenamefont
  {Zakharov}, \citenamefont {Komolkin},\ and\ \citenamefont
  {Maliniak}}]{Zakharov1999}%
  \BibitemOpen
  \bibfield  {author} {\bibinfo {author} {\bibfnamefont {A.~V.}\ \bibnamefont
  {Zakharov}}, \bibinfo {author} {\bibfnamefont {A.~V.}\ \bibnamefont
  {Komolkin}},\ and\ \bibinfo {author} {\bibfnamefont {A.}~\bibnamefont
  {Maliniak}},\ }\bibfield  {title} {\bibinfo {title} {Rotational viscosity in
  a nematic liquid crystal: A theoretical treatment and molecular dynamics
  simulation},\ }\href {https://link.aps.org/doi/10.1103/PhysRevE.59.6802}
  {\bibfield  {journal} {\bibinfo  {journal} {Phys. Rev. E}\ }\textbf {\bibinfo
  {volume} {59}},\ \bibinfo {pages} {6802} (\bibinfo {year}
  {1999})}\BibitemShut {NoStop}%
\bibitem [{\citenamefont {Conklin}\ \emph {et~al.}(2018)\citenamefont
  {Conklin}, \citenamefont {Tovkach}, \citenamefont {Vi\~nals}, \citenamefont
  {Calderer}, \citenamefont {Golovaty}, \citenamefont {Lavrentovich},\ and\
  \citenamefont {Walkington}}]{Conklin2018}%
  \BibitemOpen
  \bibfield  {author} {\bibinfo {author} {\bibfnamefont {C.}~\bibnamefont
  {Conklin}}, \bibinfo {author} {\bibfnamefont {O.~M.}\ \bibnamefont
  {Tovkach}}, \bibinfo {author} {\bibfnamefont {J.}~\bibnamefont {Vi\~nals}},
  \bibinfo {author} {\bibfnamefont {M.~C.}\ \bibnamefont {Calderer}}, \bibinfo
  {author} {\bibfnamefont {D.}~\bibnamefont {Golovaty}}, \bibinfo {author}
  {\bibfnamefont {O.~D.}\ \bibnamefont {Lavrentovich}},\ and\ \bibinfo {author}
  {\bibfnamefont {N.~J.}\ \bibnamefont {Walkington}},\ }\bibfield  {title}
  {\bibinfo {title} {Electrokinetic effects in nematic suspensions:
  Single-particle electro-osmosis and interparticle interactions},\ }\href
  {https://link.aps.org/doi/10.1103/PhysRevE.98.022703} {\bibfield  {journal}
  {\bibinfo  {journal} {Phys. Rev. E}\ }\textbf {\bibinfo {volume} {98}},\
  \bibinfo {pages} {022703} (\bibinfo {year} {2018})}\BibitemShut {NoStop}%
\bibitem [{\citenamefont {Shen}\ and\ \citenamefont {Hu}(2010)}]{Shen2010}%
  \BibitemOpen
  \bibfield  {author} {\bibinfo {author} {\bibfnamefont {S.}~\bibnamefont
  {Shen}}\ and\ \bibinfo {author} {\bibfnamefont {S.}~\bibnamefont {Hu}},\
  }\bibfield  {title} {\bibinfo {title} {A theory of flexoelectricity with
  surface effect for elastic dielectrics},\ }\href
  {https://www.sciencedirect.com/science/article/pii/S0022509610000499}
  {\bibfield  {journal} {\bibinfo  {journal} {J. Mech. Phys. Solids}\ }\textbf
  {\bibinfo {volume} {58}},\ \bibinfo {pages} {665} (\bibinfo {year}
  {2010})}\BibitemShut {NoStop}%
\end{thebibliography}%

\end{document}


\preprint{APS/123-QED}

\title{Supplemental Material for \\ ``Nematronics: Reciprocal coupling between ionic currents and nematic dynamics"}

\author{Chau Dao}
    \affiliation{Department of Physics and Astronomy and Bhaumik Institute for Theoretical Physics, University of California, Los Angeles, California 90095, USA}
\author{Jeffrey C. Everts}
    \affiliation{Institute of Physical Chemistry, Polish Academy of Sciences, 01-224 Warsaw, Poland}
    \affiliation{Institute of Theoretical Physics, Faculty of Physics, University of Warsaw, Pasteura 5, 02-093 Warsaw, Poland}
\author{Miha Ravnik}
    \affiliation{Faculty of Mathematics and Physics, University of Ljubljana, Jadranska 19, 1000 Ljubljana, Slovenia}
    \affiliation{Department of Condensed Matter Physics, Jozef Stefan Institute, Jamova 39, 1000 Ljubljana, Slovenia}
\author{Yaroslav Tserkovnyak}
    \affiliation{Department of Physics and Astronomy and Bhaumik Institute for Theoretical Physics, University of California, Los Angeles, California 90095, USA}

\maketitle

In this Supplemental Material, we provide (i) the generalized theory for nematic electrolytes with hydrodynamic effects restored, (ii) expanded discussion of the ``artificial symmetries," and (iii) proof that flexoelectric effects can be qualitatively distinguished from the dissipative nemato-ionic coupling.

\subsection{(i) Nematic hydrodynamics}
 
This model describes a uniaxial nematic electrolyte, a liquid crystal in the nematic phase doped with ions. The fluid is taken to be incompressible, that is, $\grad\cdot\*v = 0$. The electrolyte consists of freely moving monovalent ionic species. The ionic density for each species is given by $\rho^\alpha(\*r,t)$, whereby $\alpha$ indexes over each species. In this model, we will describe the simplest case of a dilute electrolyte with two ionic species consisting of a cationic species and an anionic species. However, it is straightforward to generalize and include more ionic species. $\rho^+(\*r,t)$ corresponds to the cationic density and $\rho^-(\*r,t)$ corresponds to the anionic density. We assume that the liquid crystal has uniform nematic order. Consequently, the orientational structure is described by the nematic director field $\*n(\*r,t)$, with unit norm. This model incorporates ionic degrees of freedom into the Ericksen-Leslie formalism of nematodynamics \cite{Leslie1979}. Similar models have been derived which focused on electrohydrodynamic effects but did not include our proposed reciprocal nemato-ionic coupling \cite{Tovkach2016,Tovkach2017,Calderer2016}. Henceforth, we shall adopt the convention that repeated indices are summed over unless stated otherwise. The equations of nematodynamics are derived by employing the principle of least dissipation, given by \cite{Gyarmati1970}
\begin{equation}
    \frac{\delta}{\delta\dot{q}_i}\left\{\frac{\t d \mc F}{\t d t} + \int d^3\*r \, \mc R\right\} = 0.
\end{equation}
Here $\t d/\t dt = \partial_t + \*v \cdot \bs \grad$ is the material (i.e. advective) derivative, $\mc F$ is the free energy, and $\mc R$ is our Rayleigh dissipation function. The generalized velocities are $\dot q_i$, which in our model are given by the fluid velocity $\*v$, the time derivative of the director $\t d\*n/\t dt$, and the ion current velocity for each ionic species $\*u^\alpha$. Once we specify the free energy $\mc F$ and the Rayleighan $\mc R$, the dynamic equations can be derived. First, the free energy is given by 

\begin{equation}\label{eq:free_energy}
\begin{aligned}
    \mc F[\varrho_m, \rho_\pm, \*v, \*n, \psi] = \int  d^3\*r \,\Bigg\{ &\frac{1}{2}\varrho_m \*v^2 &&\text{fluid kinetic energy}\\ +
    &\frac{K}{2}(\partial_i n_j)^2 &&\t{nematic free energy} \\ + 
    &\sum_{\alpha = \pm} k_B T\rho^\alpha\left[\ln\left(\rho^\alpha\Lambda_\alpha\right)-1\right] &&\t{nonelectrostatic ionic free energy}\\ + 
    &e(\rho_+ - \rho_-)\psi -\frac{1}{8\pi} \grad\psi \cdot \bs \epsilon(\*n) \cdot \grad \psi &&\t{electrostatic free energy}\\ +
    &\left[ g_1(\*n \cdot \grad)\*n + g_2 \*n(\grad\cdot\*n)\right]\cdot \grad\psi\Bigg\}. &&\t{flexoelectric free energy}
\end{aligned}
\end{equation}
The free energy is a functional of the local density of the fluid $\varrho_m$, the density for each ion species $\rho^\alpha$, the fluid velocity $\*v$, the director $\*n$, and the electrostatic potential $\psi$. $K$ is the Frank elastic constant describing nematic elasticity, $\Lambda_\alpha$ is the thermal de Broglie wavelength of each ionic species, $\epsilon_{ij} = \epsilon_\perp\delta_{ij} + \Delta \epsilon \, n_in_j$ is the dielectric tensor, and $g_1$ and $g_2$ are phenomenological coefficients.

The first term in Eq. \eqref{eq:free_energy} is the kinetic energy density of the fluid. The second term describes the free energy of the nematic texture, whereby we have invoked the equal constant approximation. The third term is the free energy density corresponding to the nonelectrostatic entropic free energy contributions of each ionic species. In general, this term would include the chemical interactions between each ionic species. However, when the equilibrium ion density $\rho_0$ is sufficiently small, i.e. the electrolytic solution is dilute, the free energy contribution of each ionic species can be well approximated by that of an ideal gas. The fourth term is the electrostatic contribution of the free energy corresponding to the coupling of the charge density with the electric potential, as well as the energy density of the electric field. The fifth term is the free density energy due to flexoelectricity. Here, the flexoelectric polarization 
\begin{equation}
\bs{\mc P} = g_1(\*n \cdot \grad)\*n + g_2 \*n(\grad\cdot\*n)\label{eq:flexo}
\end{equation}
is due to the strain of the nematic texture. Alternatively, it can be equivalently expressed as $\bs{\mc P} = e_1 \*n(\grad\cdot\*n) - e_3\*n\times(\grad\times\*n)$, where $e_1$ is the phenomenological coefficient corresponding to splay distortions and $e_3$ is the coefficient corresponding to bend distortions \cite{chandrasekhar_1992}.

To formulate the dynamic equations, we now specify the Rayleigh dissipation functions. The Rayleighan $\mc R$ must be quadratic in the aforementioned generalized velocities: $\dot {\*n} \equiv \t d\*n/\t dt$, $\*u^\alpha$, and $\*v$. Furthermore, it must be positive definite, and frame independent. The final constraint implies that the Rayleighan must remain invariant under a Galilean transformation and that the dissipation function must be constructed from tensors that vanish in the case of uniform fluid rotation. These tensors are given by \cite{Prost1993, Vertogen1988}

\begin{align}
    \*N &= \dot{\*n} - (\grad\times\*v)\times\*n/2, ~~~ A_{ij} = \partial_i v_j + \partial_j v_i.
\end{align}

Here, $\*N$ can be formally understood as the time derivative of the director with respect to the flow vorticity of the fluid. $A_{ij}$ is the symmetrized component of the gradient velocity field. Expressed in terms of $\*N$ and $A_{ij}$, the Rayleigh functions are given by

\begin{subequations}
    \begin{align}
        \mc R_{vv} &= \frac{1}{2}\eta_1(A_{ij})^2 + \frac{1}{2}\eta_2(A_{ij}n_j)^2\label{eq:R_vv}\\ 
        \mc R_{nv} &= \frac{1}{2}\alpha_1N_i N_i+\alpha_2N_iA_{ij}n_j,\label{eq:R_nv}\\
        \mc R_{uv} &= \frac{1}{2}k_BT \rho^\alpha (\*u^\alpha - \*v) \cdot (\bs{\mc D}^\alpha)^{-1} \cdot (\*u^\alpha - \*v),\label{eq:R_uv} \\
        \mc R_{nu} &=  \gamma^\alpha \rho^\alpha \*N \cdot\left\{ \big[(\*u^\alpha - \*v)\cdot\grad\big]\*n\right\}.\label{eq:R_nu}
    \end{align}
\end{subequations}
In the above Rayleigh functions, $\eta_1$ is the ordinary viscosity of an incompressible fluid \cite{Prost1993,Vertogen1988}. $\eta_2$ is a phenomenological viscosity coefficient that accounts for the orientation of the director. $\alpha_1$ characterizes the rotational viscosity associated with the nematic dynamics \cite{Zakharov1999}. $\alpha_2$ is a phenomenological parameter corresponding to the viscosity between nematic dynamics and the spatial gradient of the fluid velocity field. $\gamma^\alpha$ is a phenomenological parameter that characterizes the strength of the reciprocal nemato-ionic coupling. $\bs{\mc D}^\alpha(\*r,t)$ is the diffusion tensor corresponding to each ionic species, constructed on general symmetry grounds as $\mc D^{\alpha}_{ij} = \mc D^{\alpha}_{\perp} \delta_{ij} + \Delta \mc D^\alpha\,n_in_j$. $\mc R_{vv}$ describes the dissipation due to the fluid viscosity, $\mc R_{nv}$ describes the friction between the nematic dynamics and the fluid. $\mc R_{uv}$ describes the friction between the ionic currents $\*u^\alpha$ and the background fluid flow given by $\*v$. Eqs. (\ref{eq:R_vv} -- \ref{eq:R_uv}) appeared in Refs. \cite{Tovkach2016,Tovkach2017,Conklin2018}, which discussed the electrohydrodynamic effects that arise due to these Rayleigh functions. $\mc R_{nu}$ describes the friction between the nematic dynamics and ionic currents. 

Having specified the free energy $\mc F$ and the Rayleighan $\mc R$, we now employ the principle of least dissipation to derive the equations of motion. Firstly, taking the time derivative of the free energy, we have 

\begin{align}\label{eq:dtF}
    \frac{\t d\mc F}{\t dt} = \int d^3\*r \left\{\varrho_m v_i\dot{v}_i + \dot{\*n}\cdot \*h_\perp + (\*u^\alpha - \*v)\cdot\grad\mu^\alpha - \rho^\alpha\mu^\alpha (\grad\cdot\*v)+ \sigma_{ij}(\partial_iv_j)\right\}.
\end{align}
We have expressed the above equation in terms of $\*h_\perp$ and $\mu^\alpha$. $\*h_\perp = \*n \times (\delta\mc F/\delta\*n)\times\*n$ is the molecular field thermodynamically conjugate to $\*n$, with components parallel to $\*n$ projected out to fix $|\*n|=1$, and $\mu^\alpha=\delta\mc F/\delta \rho^\alpha$ is identified as the electrochemical potential thermodynamically conjugate to each ionic species. The first term is the time derivative of the kinetic energy associated with the fluid. The second, third, and fourth terms stem from the variation of the orientation of the director and the ionic density of a fluid element. To derive the third and fourth terms, we have invoked the continuity equation, $\partial_t \rho^\alpha = \grad \cdot (\rho^\alpha\*u^\alpha)$ and integrated by parts. The fifth term in the equation is the change in the free energy when the director orientation and ionic density of the fluid element are constant, but there is a material flow. $\sigma_{ij}$ can be formally understood as the stress tensor. We note that in the minimal system with quenched fluid dynamics described in the main text, only the second and third terms survive.

The following derivation of the stress tensor $\sigma_{ij}$ closely follows that given by de Gennes and Prost \cite{Prost1993}. Let us consider a generic free energy density $F(\rho^\alpha,\*n,\psi)$. A material flow corresponds to a transformation of $\*r$, $\*n(\*r)$, $\rho^\alpha(\*r)$, and $\psi$ such that

\begin{align}
   \*r \rightarrow \*r^\prime = \*r + \delta\*w(\*r), ~~~ \*n(\*r) \rightarrow \*n^\prime(\*r^\prime) = \*n(\*r + \delta \*w(\*r)), \label{eq:coord_trans}
\end{align}
where $\delta \*w(\*r)$ is the displacement of the fluid element due to the flow. $\rho^\alpha$ and $\psi$ transform similarly to $\*n$. This displacement leaves the orientation of the director and the ionic density and electric potential of the fluid element unchanged. Making the substitution $\delta \*w(\*r) \rightarrow \*v(\*r)\,dt$ will give us the rate of change of the free energy due to a material flow. To keep our equations algebraically simple, let us first calculate the stress tensor for a free energy density that depends solely on $\*n$. This case can then be easily generalized to the free energy density depending on $\*n$, $\rho^\alpha$, and $\psi$. Firstly, by expanding the transformed director field to leading order in $\delta\*w(\*r)$, we have
\begin{equation}\label{eq:trans_n}
    \*n(\*r^\prime) \approx \*n(\*r) - \delta\*w(\*r)\cdot\frac{\partial}{\partial\*r^\prime}\*n(\*r^\prime)\bigg\vert_{\*r^\prime = \*r}.
\end{equation}
The expansion for $\rho^\alpha$ and $\psi$ will be analogous. Utilizing Eq. (\ref{eq:coord_trans}), the derivative $\partial_{\*r^\prime}$ can be written as
\begin{equation}\label{eq:trans_der}
    \frac{\partial}{\partial r^\prime_i} = \frac{\partial r_j}{\partial r^\prime_i}\frac{\partial}{\partial r_j} = \Big[\delta_{ij} - \partial_i(\delta w_j)\Big]\partial_j = \partial_i - \left[\partial_i(\delta w_j)\right]\partial_j.
\end{equation}
Applying the form of the derivative in Eq. (\ref{eq:trans_der}) to Eq. (\ref{eq:trans_n}), and isolating the second term, the variation of the director field $\delta \*n(\*r)$ is
\begin{equation} 
    \delta\*n(\*r)= -\big[\delta\*w(\*r) \cdot \grad\big]\*n(\*r).
\end{equation}
Now we consider the variation of the free energy $\mc F[\*n]$ and show how it can be written in terms of the stress tensor. We have that 

\begin{subequations}
\begin{align}
    \delta \mc F[\*n] &= \int d^3\*r \left\{\frac{\partial F}{\partial n_\mu}\delta n_\mu +  \frac{\partial F}{\partial(\partial_i n_\mu)}\delta(\partial_i n_\mu)\right\} \\
    &= \int d^3\*r\left\{\left[\partial_i \left(\frac{\partial F}{\partial(\partial_i n_\mu)}\right)\right]\delta n_\mu+ \left(\frac{\partial F}{\partial(\partial_i n_\mu)}\right)\Big(\partial^\prime_i n^\prime_\mu - \partial_i n_\mu\Big)\right\}\\
    &= \int d^3\*r\left\{\left(\frac{\partial F}{\partial(\partial_i n_\mu)}\right)\left[-\partial_i(\delta n_\mu) + \partial^\prime_i(n_\mu + \delta n_\mu) - \partial_i n_\mu\right]\right\},
\end{align}
where $F$ is the free energy density. Substituting in the transformed derivative given in Eq. \ref{eq:trans_der}, and eliminating terms second order in variations, we have
\begin{align}
    \delta \mc F[\*n] &= -\int d^3\*r\left\{ \left(\frac{\partial F}{\partial(\partial_i n_\mu)}\right)\Big[\partial_i(\delta w_j)\partial_j n_\mu\Big]\right\}= \int d^3\*r\,\sigma_{ij}\partial_i(\delta w_j)
\end{align}
\end{subequations}
The stress tensor when the free energy is solely dependent on $\*n$ is given by 
\begin{equation}
    \sigma_{ij}(\*n) = -\left(\frac{\partial F}{\partial(\partial_i n_\mu)}\right)(\partial_j n_\mu).
\end{equation}
Now we generalize to the case when the free energy density is a function of $\*n$, $\rho^\alpha$, and $\psi$. The stress tensor in Eq. (\ref{eq:dtF}) thus becomes 

\begin{equation}
    \sigma_{ij}(\*n,\rho^\alpha,\psi) = \left(\frac{\partial F}{\partial(\partial_i n_\mu)}\right)(\partial_j n_\mu) + \left(\frac{\partial F}{\partial(\partial_i \psi)}\right)(\partial_j \psi) + \left(\frac{\partial F}{\partial(\partial_i \rho^\alpha)}\right)(\partial_j \rho^\alpha).
\end{equation}
Upon substitution of the free energy density given in Eq. (\ref{eq:free_energy}), this yields
\begin{equation}
    \sigma_{ij} = -K(\partial_in_k)(\partial_jn_k) + \left(\mc P_i + \frac{1}{4\pi} \epsilon_{ik}E_k\right)E_j
\end{equation}
The first term in $\sigma_{ij}$ corresponds to the elastic stress tensor while the second term corresponds to the Maxwell stress tensor \cite{Tovkach2017,Shen2010}. Having specified the Rayleighans and calculated the time derivative of the free energy, we can now perform the variations with respect to $\dot{\*n}$, $\*u^\alpha$, and $\* v$. Since there are three generalized velocities, this will yield three primary governing equations. Firstly, by varying $\dot{\mc F} + \int d^3\*r\,\mc R$ with respect to $\*u$ we get 
\begin{subequations}
    \begin{align}
    \*u^\alpha(\*r,t) &= \*v - \frac{1}{k_{B}T}\bs{\mc D}^\alpha \cdot\left( \grad\mu^\alpha + \bs{\mc E}^\alpha \right),\label{eq:ion_current}\\ \intertext{and the motive force for each ionic species $\bs{\mc E}_\alpha$ is given by}
    \mc E^\alpha_i &= \gamma^\alpha \*N\cdot \partial_i \*n \label{eq:motive_force}.
    \end{align}
\end{subequations}
Next the variation with respect to $\*v$ yields

\begin{align}
    \varrho_m \frac{\text{d}\*v}{\text dt} + \grad \cdot \left(-p\*1-\bs{\sigma} -  \frac{\partial \mc R}{\partial(\partial_i v_j)}\right) = 0
\end{align}
where $p$ is the total pressure that has absorbed all terms that are proportional to $\delta_{ij}$. It accounts for both the pressure due to regular incompressible fluid, as well as the osmotic pressure stemming from the ionic subsystem. To eliminate terms stemming from $\partial \mc R/\partial \*v$ and $\partial \dot{\mc F}/\partial \*v$, we utilize the expression for the ion current given in Eq. (\ref{eq:ion_current}). The remaining term $\partial \mc R/\partial(\partial_i v_j)$ can be understood as the ``viscous" component of the stress tensor, as opposed to the elastic and electrostatic component characterized by $\bs{\mc \sigma}$. Finally, variation with respect to $\dot{\*n}$ will yield the equation of the nematic director dynamics
\begin{equation}
    h_{\perp,i} + \alpha_1 N_i + \alpha_2A_{ij}n_j + h_{\tau,i} = 0,
\end{equation}
where $\*h_\tau$ is the dissipative field exerted by the ionic currents onto the nematic texture. It can be given by 
\begin{equation}\label{eq:h_tau}
    h_{\tau,i} = \gamma^\alpha\rho^\alpha[(\*u^\alpha-\*v)\cdot\grad]n_i,
\end{equation}
where we have summed up the contributions from each ionic species. The motive force in Eq. (\ref{eq:motive_force}) and the dissipative field in Eq. (\ref{eq:h_tau}) satisfy Onsager reciprocity. When fluid motion is quenched and we focus on the charge dynamics, we recover the expressions found in the main text for $\bs{\mc E}$ and $\*h_\tau$. 
\newpage

\subsection{(ii) Artificial symmetries | Reduced material constants}
In the main text, the free energy corresponding to the nematic texture is given by 
\begin{equation}
    \mc F_\t{artificial} = \int d^3\*r \left\{\frac{K}{2}(\partial_in_j)^2\right\},
\end{equation}
where $K$ is the Frank constant. Here, we have imposed an ``artificial" symmetry in addition to the actual symmetry of the system. This free energy density is invariant under independent rotations of real space and nematic director orientation. In the most general case, this artificial symmetry is not present, and the actual free energy is given instead by 

\begin{align}
    \mc F_\t{actual} = \int d^3\*r\left\{\frac{K_1}{2}\left(\grad \cdot \*n\right)^2 + \frac{K_2}{2}\left[\*n \cdot (\grad \times \*n)\right]^2 + \frac{K_3}{2}\left[\*n \times (\grad \times \*n)\right]^2\right\}.
\end{align}
$K_1$, $K_2$, and $K_3$ are phenomenological coefficients called the Frank elastic constants and they correspond to the splay mode, bend mode, and twist mode, respectively. These constants are generically independent parameters. We can restore the artificial symmetry by setting $K_1 = K_2 = K_3 = K$, and this treatment is commonly referred to as the ``equal-constant approximation" which simplifies calculations but still gives good qualitative insights into the system. Within a fluid, the remaining symmetry that the system has is simultaneous uniform rotations of real space and order parameter space. This means that when hydrodynamics is restored, the Rayleigh dissipation function must vanish under a rigid fluid rotation. Consequently, the Rayleigh functions must be constructed out of tensors that appropriately vanish under a rigid rotation. Instead of $\dot{\*n}$ the relevant tensor is now 
\begin{equation}
    \*N = \dot{\*n} - (\grad \times \*v)\times \*n/2,
\end{equation}
which reduces to $\partial_t\*n$ when the fluid motion is quenched. In the construction of $\mc R_{nj} = \gamma\, \partial_t\*n\cdot [(\*j \cdot \grad) \*n]$ in the main text, the charge current $\*j$ was coupled to $\partial_t\*n$ through the gradient of the nematic texture, $\grad \*n$ to maintain the artificial symmetry. Substituting in $\*N$ for $\partial_t\*n$ in $\mc R_{nj}$, we get the Rayleigh function given in Eq. (\ref{eq:R_nu}). Because of the dependence of $\*N$ on $\*v$, $\mc R_{nj}$ is no longer independently isotropic in real space as well as order parameter space. 

Even in the absence of fluid motion, there can be additional terms that break this artificial symmetry. For simplicity, let us retain only the terms that are relevant in the case of a uniform charge current $\*j$. With this constraint, the effective dissipative field $\*h_\tau$ and the corresponding motive force $\bs{\mc E}$ can be constructed as follows:
\begin{subequations}
    \begin{align}
        h_{\tau,i} &= \gamma_1 (\*j\cdot\grad)n_i + \gamma_2\, \*j\cdot \partial_i\*n + \gamma_3 \, j_i (\grad\cdot\*n), \\
        \mc E_i &= \gamma_1 (\partial_t \*n)\cdot(\partial_i\*n) + \gamma_2 (\partial_t\*n\cdot\grad)n_i + \gamma_3 (\partial_tn_i)(\grad\cdot\*n).
    \end{align}
\end{subequations}
Here, the only symmetry is that of a combined rotation of order parameter space and real space.

\newpage

\subsection{(iii) Qualitatively distinguishing flexoelectric effects}

In the illustrative examples provided in the main text, we did not include terms in the free energy, Eq. (2) which stem from flexoelectric effects. Here, we discuss how flexoelectric effects can be qualitatively distinguished from those due to the nemato-ionic coupling in each of the examples. When the system size is much larger than the ionic screening length, electrostatic effects as well as flexoelectric effects are screened. Dynamic effects due to the nemato-ionic coupling will subsequently dominate. On the other hand, when the screening length is large (e.g. at low ion concentrations), we must consider the linear coupling of the electric field with the flexoelectric polarization, given by the free energy density $F_f = -\bs{\mc P}\cdot \*E$. Since the flexoelectric free energy density is linear in $\*E$, flexoelectricity will affect director dynamics in the linear response theory in addition to the dissipative field $\*h_\tau$. To see how these effects can be qualitatively distinguished from those which we focus on, let us examine the dynamics of the bound charge induced by flexoelectricity in the context of each of our examples. 

In the setup of the first example of the hybrid aligned nematic cell, the polarization induces a bound charge density $\rho^b = -\grad \cdot \bs{\mc P}$.  When no current is applied, the director in the ground state will undergo uniform $\pi/2$ winding, and the bound charge density of this texture is
\begin{equation}
    \rho^b(z) = -\frac{\pi}{2d}(g_1 + g_2)\cos(\pi z/d).
\end{equation}
The bound charge density is antisymmetric about $z=d/2$. Applying an $\*E$ field would cause the positively and negatively charged regions to either draw together or pull apart symmetrically. This corresponds to winding density symmetrically localizing at the center or edges, respectively. Consequently, this is qualitatively different than the response of the nematic to the charge current, which would result in the winding asymmetrically ``bunching up" at the edges. 

We now discuss the second example. In the case of the line disclination subject to a current flow, the dissipative field $\*h_\tau$ exerted by the charge current onto the nematic texture is agnostic to the winding number. This is because both the Rayleigh dissipation function and the free energy are invariant to a global $\pm\pi$ rotation of the director field about the $z$ axis, which would change the $+1/2$ defect to a $-1/2$ defect. On the other hand, the bound charge
\begin{equation}
    Q^b_{\pm 1/2} = \pm\pi(g_2 - g_1)
\end{equation}
enclosed in the vicinity of the defect changes sign based on the winding number. Whereas the dissipative field would always drag the soliton in the direction of the applied current, flexoelectric effects dictate that a $+1/2$ and a $-1/2$ defect move in opposite directions when subject to an external electric field. The asymmetric characteristic in the hybrid aligned nematic cell and the sign change of the bound charge enclosed in the vicinity of the defect provide the distinction between the flexoelectric and nemato-ionic effects.

\bibliography{supplement.bib}